\documentclass[fleqn,usenatbib,useAMS]{mnras}

\usepackage{newtxtext,newtxmath}

\usepackage[T1]{fontenc}
\usepackage{ae,aecompl}

\DeclareRobustCommand{\VAN}[3]{#2}
\let\VANthebibliography\thebibliography
\def\thebibliography{\DeclareRobustCommand{\VAN}[3]{##3}\VANthebibliography}

\usepackage{graphicx}	
\usepackage{amsmath}	

\usepackage{amssymb}	
\usepackage{array}
\usepackage{multirow}
\usepackage{tabularray}
\usepackage{subcaption} 
\usepackage{multirow}
\usepackage{hyperref}
\usepackage{multicol}        
\usepackage{bm}		
\usepackage{pdflscape}	

\newcommand{\angstrom}{\textup{\AA}}

\title[High-$z$ quasar candidate archive]{High-$z$ Quasar Candidate Archive: A Spectroscopic Catalog of Quasars and Contaminants in Various Quasar Searches}

\author[Yang et al.]{
Da-Ming Yang,$^{1}$\thanks{E-mail: dyang@strw.leidenuniv.nl}
Jan-Torge Schindler,$^{1,3}$
Riccardo Nanni,$^{1,2}$
Joseph F. Hennawi,$^{1,2}$
Eduardo Ba\~nados,$^{3}$
\newauthor{
Xiaohui Fan,$^{4}$
Anniek Gloudemans,$^{1}$
Huub Rottgering,$^{1}$
Feige Wang,$^{4}$
and Jinyi Yang$^{4}$}
\\
$^{1}$Leiden Observatory, Leiden University, P.O. Box 9513, 2300 RA Leiden,
the Netherlands\\
$^{2}$Department of Physics, Broida Hall, University of California, Santa Barbara, Santa Barbara, CA 93106-9530, USA\\
$^{3}$Max-Plank-Institut für Astronomie, Königstuhl 17, D-69117 Heidelberg, Germany\\
$^{4}$Steward Observatory, University of Arizona, 933 North Cherry Avenue, Tucson, AZ 85721, USA
}

\date{Accepted XXX. Received YYY; in original form ZZZ}

\pubyear{2022}

\begin{document}
\label{firstpage}
\pagerange{\pageref{firstpage}--\pageref{lastpage}}
\maketitle

\begin{abstract}
We present the high-$z$ quasar candidate archive (HzQCA), summarizing the spectroscopic observations of 174 $z\gtrsim5$ quasar candidates using Keck/LRIS, Keck/MOSFIRE, and Keck/NIRES. We identify 7 candidates as $z\sim 6$ quasars 3 of them newly reported here, and 51 candidates as brown dwarfs. In the remaining sources, 74 candidates are unlikely to be quasars; 2 sources are inconclusive; the others could not be fully reduced or extracted. Based on the classifications we investigate the distributions of quasars and contaminants in color space with photometry measurements from DELS ($z$), VIKING/UKIDSS ($YJHK_s$/$YJHK$), and un\textit{WISE} ($W1W2$). We find that the identified brown dwarfs are not fully consistent with the empirical brown dwarf model that is commonly used in quasar candidate selection methods. To refine spectroscopic confirmation strategies, we simulate synthetic spectroscopy of high-$z$ quasars and contaminants for all three instruments. The simulations utilize the spectroscopic data in HzQCA. We predict the required exposure times for quasar confirmation and propose and optimal strategy for spectroscopic follow-up observations. For example, we demonstrate that we can identify a $m_J=21.5$ at $z=7.6$ or a $m_J=23.0$ at $z=7.0$ within 15\,min of exposure time with LRIS. With the publication of the HzQCA we aim to provide guidance for future quasar surveys and candidate classification.
\end{abstract}




\section{Introduction}\label{sec:intro}

Large samples of high-$z$ quasars can help us understand various fundamental questions in extragalactic astronomy. The existence of quasars with ${\rm \sim 10^9\ M_\odot}$ super massive black hole (SMBH) in their centers only ${\rm \sim 1\ Gyr}$ after the big bang already poses a stringent constrain on the classic SMBH formation theory \citep[e.g.][]{Volonteri2010}. Theories involve the formation of massive black hole seeds ($M>10^4 M_\odot$) via direct collapse \citep[e.g.][]{Begelman2006,Dayal2019}, or rapid growth through super-Eddington and hyper-Eddington accretion \citep[e.g.][]{Inayoshi2016,Davies2019} have been proposed to explain the observations \citep[see][for a recent review]{Inayoshi2020}. Further estimation of the accretion timescale of certain quasars, i.e. the quasar lifetime $t_{\rm Q}$, with the proximity zones in their spectra posts even more challenges in the accretion theory \citep[e.g.,][]{Eilers2017,2018ApJ...867...30E,2020ApJ...900...37E}. Moreover, high-$z$ quasar is a powerful tool to study the evolution of the epoch of reionization (EoR). The damping wing feature \citep{1998ApJ...501...15M} in spectra of high-$z$ quasars implies that the intergalactic medium (IGM) is significantly neutral at $z\gtrsim 7$ \citep{Mortlock2011,Banados2018,Davies2018,Yang2020A,Wang2021} but highly ionized at $z\lesssim 6$ \citep[e.g.][]{Eilers2018,Yang2020B}. Such measurements provide an important way to probe the evolution of the neutral fraction during the EoR, complementing the results based on the cosmic microwave background (CMB) polarization \citep{PlanckVII}.

Since the discovery of the first quasar (QSO) 3C273 at $z=0.158$ \citep[][]{Schmidt1963}, astronomers have invested tremendous effort in finding quasars with increasingly higher redshifts.  Because quasars are rare on the sky, these searches are typically conducted using large imaging surveys.
Optical surveys such as SDSS \citep[e.g.,][]{Fan2006,Jiang2008}, Pan-STARRS \citep[e.g.,][Ba\~nados et al. in prep.]{Banados2016B}, DELS \citep[e.g.,][]{Wang2017,Wang2019}, and SHELLQs \citep[e.g.,][]{Matsuoka2022} have been extensively searched for quasars in the range $5.7<z\leq 7.0$. To date, over 400 quasars with $z>5.7$ have been found, with $\sim 70$ of them have $6.5<z<7$ \citep[a list of all $z>5.7$ quasars with frequent updates can be found in ][]{Bosman2020}. Several searches based on near-infrared (NIR) imaging surveys have also paved the way for the 
discovery of quasars with even higher redshifts, such as the United Kingdom Infrared Telescope (UKIRT) Infrared Deep Sky Survey \citep[UKIDSS; e.g.][]{Mortlock2011,Banados2018}, the Visible and Infrared Survey Telescope for Astronomy (VISTA) Kilo-degree Infrared Galaxy Survey \citep[VIKING; e.g.][]{Venemans2013}, and the UKIRT Hemisphere Survey \citep[UHS; e.g.][]{Yang2020A,Wang2021}. With the IR surveys, three quasars with $z>7.5$ \citep{Banados2018,Yang2020A,Wang2021} have been found in the past few years.

The study of high-$z$ quasars is about to be revolutionized by the advent of the \textit{Euclid} mission \citep[][]{Laureijs2011,Barnett2019}. With the $OYJH$ multi-band photometry\footnote{The 5$\sigma$ depths of $OYJH$ are $m=24.5$, $m=24.0$, $m=24.0$, and $m=24.0$, respectively.} and the 15,000 ${\rm deg}^2$ sky coverage provided by \textit{Euclid}, we expect to find over 100 quasars with $7.0 < z < 7.5$, and $\sim 25$ quasars beyond $z = 7.5$, including $\sim 8$ beyond $z = 8.0$. However, quasar searches with \textit{Euclid} also introduce several new challenges. The first one is a requirement for a more efficient candidate selection algorithm. 
The expected number density of $z\gtrsim 7$ quasars is $\sim 10^{-3}\ {\rm deg}^2$ at $J=21$ \citep[][]{Wang2019}, while the contaminants, mostly Galactic brown dwarfs, are far more numerous, $\sim 10\ {\rm deg}^2$ at $J=21$ \citep[][]{Barnett2019}. Meanwhile, as \textit{Euclid} has a deeper flux limit, there will be a large number of faint candidates. However, the widely used color selection methods will be less efficient when selecting candidates from faint sources owing to the low quality of the photometry measurements. The low success rate makes this kind of selection method not suitable for quasar search with \textit{Euclid} with a large number of candidates. More advanced probabilistic selection methods have been proposed to segregate quasars from the contaminants in an attempt to improve the success rate \citep[e.g.][]{Mortlock2012,Bovy2011,Nanni2022}. With the method in \cite{Mortlock2012}, several successful quasar searches with optical survey have been conducted \citep[e.g.][]{Matsuoka2022}, leading to an abundant discoveries of $z<7$ quasars with very high success rate. \cite{Nanni2022} proposed the \texttt{XDHZQSO}, another Bayesian approach based on extreme deconvolution \citep[XD;][]{Bovy2011}. The model for contaminants is purely empirical and all the contaminants are considered as one class. These probabilistic methods need further refinement to prepare for future surveys, and the key to do so is a deeper understanding towards the contaminant population. Moreover, the regime of higher redshift and/or fainter luminosity is poorly explored. Quasars that can be found by \textit{Euclid} will be as faint as $m_J>23.0$ \citep[][]{Barnett2019}, whereas we have only found very few quasars with $m_J>22.0$. In addition, only 9 quasars with $z>7$ have be discovered \citep[][]{Mortlock2011,Banados2018,Wang2018,Yang2019,Matsuoka2019A,Matsuoka2019B,Yang2020A,Wang2021}, while with \textit{Euclid} we expect to find over 100 of them as aforementioned. Before stepping into this unexplored regime, we have to answer the following questions: Can we still use ground-based telescopes to effectively confirm a considerable number of quasars with higher redshifts or fainter magnitudes? If so, what is the optimal observing strategy, i.e. for quasars with different redshifts, which instruments provide the fastest confirmation and what are the required exposure times?

To cope with these difficulties, in this paper we exploit the spectroscopic observations of hundreds of high-$z$ quasar candidates with Keck in the past three years. We reduce all the spectra in a consistent manner and construct a catalog of all of the observations, which we refer to as the High-$z$ Quasar Candidate Archive (HzQCA). Using this archive, we explore both the contaminants in our candidate samples and the optimal observing strategies to provide insights for future quasar searches. First, we investigate the distribution of the contaminants in color-space, and compare it with the brown dwarf model adopted in many quasar selection methods. Second, we use real archival data to 
simulate spectroscopic observations of mock quasars, allowing us to determine the exposure times required to identify \textit{Euclid} quasars with Keck instruments.   The results also aid us to establish an optimal spectroscopic confirmation strategy.

In Section \ref{sec:archive}, we describe the construction of the archive, including the surveys involved, the instruments we use for confirmation, and the data reduction procedure. In Section \ref{sec:result}, we present the final data products and analyze the contaminant population. In Section \ref{sec:refine}, we introduce the simulations used to estimate the required exposure times and discuss refinements to the observing strategy. Throughout the paper, we adopt a flat cosmological model with $H_0=68.5\ {\rm km\ s^{-1}\ Mpc^{-1}}$, $\Omega_{\rm M}=0.3$, $\Omega_{\rm \Lambda}=0.7$. All the magnitudes are given in the AB system, while the uncertainties of our reported measurements are at $1\sigma$ confidence level.

\section{Construction of the archive}\label{sec:archive}

The HzQCA includes all the high-$z$ quasar candidates observed in two Keck programs. The first one, \textit{Searching for Quasars Deep Into the Epoch of Reionization} (U170, PI: Hennawi) uses the Near-InfraRed Echellette Spectrometer \citep[NIRES;][]{2004SPIE.5492.1295W} and the Multi-Object Spectrometer For Infra-Red Exploration \citep[MOSFIRE;][]{2012SPIE.8446E..0JM} to hunt for $z>7$ quasars from 2018 to 2021. The second one, \textit{Paving the Way for Euclid and JWST via Optimal Selection of High-z Quasars} (U055, PI: Hennawi) uses the Low Resolution Imaging Spectrometer \citep[LRIS;][]{1995PASP..107..375O,2010SPIE.7735E..0RR} and MOSFIRE to search for high-$z$ quasars with the \texttt{XDHZQSO} algorithm. The latter program is designed to demonstrate, characterize and improve the efficacy of the \texttt{XDHZQSO} in preparation for quasars selection with \textit{Euclid} and \textit{JWST}. These two programs comprise a total of 20 nights in 10 observing runs, as listed in Table \ref{tab:obs_run}. The targets in these runs were observed with a typical seeing of 0\farcs5-1\farcs2. During the observations, on-the-fly `quick-look' reductions were performed with \texttt{PypeIt} (see Section \ref{sec:reduction}) allowing
the observers to classify the candidates in real time. We usually terminated sequences if there was no sign of Lyman-break in the quick-look results, which is part of our observing strategy.

\begin{table*}
	\centering
	\caption{Observation runs included in the archive. The `Surveys' column indicates the quasar surveys involved in the given observing runs, see the description in Section \ref{sec:archive}. `Number of Candidates' means the number of candidates of which we took spectroscopic data during the observation. One source was observed in three different observing runs. Therefore, the total number in the table refers to the number of candidates.}
	\label{tab:obs_run}
	\begin{tabular}{llclcc}
		\hline
		Observation Run & Dates & Keck Program ID & Surveys &  Seeing & Number of Candidates \\
		\hline
		NIRES-1903 & March 03-04, 2019 & U170 & Z7DROPOUT & 0\farcs7-1\farcs1 & 15  \\
		NIRES-1905 & May 19, 2019 & U170 & Z7DROPOUT & 0\farcs6-1\farcs0 & 6 \\
		MOSFIRE-1911 & November 18-20, 2019 & U179 & Z7DROPOUT & 0\farcs9-1\farcs2 & 13 \\
		MOSFIRE-2005 & May 27-29, 2020 & U179 & Z7DROPOUT & 0\farcs5-0\farcs7 & 13  \\
		MOSFIRE-2010 & May 22-25, 2020 & U179 & Z7DROPOUT & 0\farcs5-0\farcs7 & 34 \\
		MOSFIRE-2201 & January 10-11, 2022 & U055 & XDHZQSO, Z7DROPOUT & 0\farcs6-0\farcs7 & 7 \\
		MOSFIRE-2204 & April 9, 2022 & U055 & XDHZQSO, Z7DROPOUT & 0\farcs6-1\farcs2 & 8 \\
		LRIS-2201 & January 27, 2022 & U055 & XDHZQSO, PS1COLOR & 0\farcs8-1\farcs0 & 19 \\
		LRIS-2203 & March 05-06, 2022 & U055 & XDHZQSO, PS1COLOR & 0\farcs6-0\farcs8 & 31 \\
		LRIS-2204 & April 23, 2022 & U055 & XDHZQSO, LOFAR, Z7DROPOUT & 0\farcs8-1\farcs0 & 30 \\
		\hline
		Total & & &  & & 174  \\
		\hline
	\end{tabular}
\end{table*}

In Section \ref{sec:survey}, we summarize the different quasar searches that provide the candidates. In Section \ref{sec:spectroscopy}, we give a brief introduction of the spectrometers used in this work. In Section \ref{sec:reduction}, we describe our data reduction procedures. In Section \ref{sec:classification}, we present our classification scheme to classify the candidates with their reduced spectra.

\subsection{Quasar Searches}\label{sec:survey}

The quasar candidates observed in the aforementioned observing runs comprise sources from various different quasar searches. We briefly summarize
them in this section. For brevity, we use short names of the selection methods to represent different searches. The imaging surveys involved are introduced in each subsection of the search.

\subsubsection{XDHZQSO: VIKING/UKIDSS+WISE+DELS+PS1}\label{sec:VIK+UKI}

\cite{Nanni2022} searched for high-$z$ quasars in the $\sim$1,000 $\ {\rm deg}^2$ overlapping area from the DESI Legacy Imaging Surveys \citep[DELS;][]{Dey2019}\footnote{\url{https://www.legacysurvey.org/}}, the VISTA Kilo-degree Infrared Galaxy (VIKING) Survey \citep[][]{Edge2013}, and the un\textit{WISE} imaging surveys \citep[][]{Lang2014} with the \texttt{XDHZQSO} method. They studied the distributions of both quasars and contaminants in the color space constructed by $z$-band from DELS, $YJHK_s$ from VIKING, and $W1W2$ from un\textit{WISE}\footnote{They performed forced photometry on VIKING images with a 1\farcs5 aperture. For sources covered by the DELS but not detected, they performed forced photometry on the Dark Energy Camera Legacy Survey (DECaLS) images with a 1\farcs5 aperture and on the un\textit{WISE} images with a 7\arcsec aperture.}. Specifically, they trained the probability density of the contaminants on 1,902,071 observed sources covered by the aforementioned surveys, and the probability density of quasars on simulated sources with synthetic photometry from \texttt{simqso} \citep{McGreer2013}. With the trained densities, they predicted the probability of a given source being a quasar in certain redshift bin ($P_{\rm QSO}$), and sources with $P_{\rm QSO}>0.1$ were selected as candidates. There were 138, and 43 candidates in the range $6\leq z\leq 7$, and $7\leq z\leq 8$, respectively. See \cite{Nanni2022} for more detailed descriptions of the methodology and the survey. 

Following the approach from \cite{Nanni2022}, we performed another search in the $\sim 4,000\ {\rm deg}^2$ UKIRT Infrared Deep Sky Survey \citep[UKIDSS;][]{Lawrence2007} footprint. At NIR wavelengths, we used $Y$-, $J$-, $H$-, and $K$-bands from UKIDSS DR11.
The UKIDSS data were obtained from the UKIRT Science Archive\footnote{\url{http://horus.roe.ac.uk/wsa/}}. We also used the $W1$- and $W2$-bands from the un\textit{WISE} release. For optical bands, we used $grz$-bands data from the DELS, and the $grizy$-bands data from Pan-STARRS \citep[PS1;][]{Chambers2016} in our selection. The detailed procedures to construct the UKIDSS candidate list are as follows:

\begin{enumerate}
    \item We selected all the sources with $J$-band signal-to-noise ratio ${\rm S/N}(J)\ge5$ first, amounting for a total of 52,273,874 sources. We also removed bright sources ($m_J<17$), as we found they were often artifacts or bright stars, after performing a visual inspection of a few hundreds of them.
    \item Since the UKIDSS survey is affected by the presence of a large number of spurious sources resulting from detector cross-talk \citep{Dye2006}, we decided to split the UKIDSS $J$-band detected sample into two sub-catalogs: one used for $6\le z \le 7$ QSO selection and another for $7\le z \le 8$ QSO selection.
    \item We constructed the $6\le z \le 7$ QSO search catalog by cross-matching the UKIDSS $J$-band detected sources with DELS, using a radius of 2\arcsec. We only kept sources with ${\rm S/N}(z)\ge 5$. Since $z\ge 6$ QSOs drop out in the bluest optical filters, we further required our objects to have ${\rm S/N}(g,r)<3$.
    \item On the other hand, we constructed the $7\le z \le 8$ QSO search catalog by cross-matching the UKIDSS $J$-band detected sources with un\textit{WISE}, using a radius 2\arcsec. We kept all the sources with ${\rm S/N}(W1)\ge 5$. For these sources we performed forced photometry on the Dark Energy Camera Legacy Survey \citep[DECaLS;][]{Dey2019} images with an aperture radius of 1\farcs5, and then further required them to have ${\rm S/N}(g,r)<3$.
    \item Then, for both the $6\le z \le 7$ and $7\le z \le 8$ QSO search catalogs, we performed forced photometry on the PS1-$grizy$ band images and removed objects with ${\rm S/N}(g_{\rm PS1},r_{\rm PS1})\ge 3$, and ${\rm S/N}(i_{\rm PS1})\ge5$ and $i_{\rm PS1}-z_{\rm PS1}<2$.
    \item For the surviving sources, we performed forced photometry on the UKIDSS-$YHK$ images, using an aperture radius of 1\farcs5.
    \item Finally, we removed sources that have no coverage in all the requested filters (UKIDSS-$YHK$, DECaLS-$z$, PS1-$izy$, un\textit{WISE}-$W1W2$). The final two parent catalogs contains 204,882, and 458,513 sources at $6\le z \le 7$ and $7\le z \le 8$, respectively.
\end{enumerate}
        
We used these two UKIDSS parent catalogs, appropriate for selecting either $6\le z \le 7$ or  $7\le z \le 8$ quasars, as the inputs
to construct the  contaminant models in the \texttt{XDHZQSO} Bayesian formalism. Adopting the same quasar probability threshold used by \cite{Nanni2022}, $P_{\rm QSO}>0.1$, we obtained a total of 48, and 83 candidates at $6\le z \le 7$ and $7\le z \le 8$, respectively. 
To date, we have observed 78 sources from all the candidates in the VIKING/UKIDSS surveys. Among them, 29, 9 are $6\leq z\leq7$, $7\leq z\leq 8$ candidates from UKIDSS, and 40, 10 of them are $6\leq z\leq 7$, $7\leq z\leq 8$ candidates from VIKING, respectively. Ten sources were selected by both searches.

\subsubsection{Z7DROPOUT: UKIDSS/UHS/VHS+WISE+DELS+PS1}\label{sec:Z7DROPOUT}

A series of searches following the selections of \cite{Banados2018}, \cite{Yang2020A}, and \cite{Wang2021} was conducted by the same investigators. These searches aimed at finding quasars with $z>7.2$ ($z$-band dropout) or even $z>7.5$ ($y$-band dropout) among sources that dropped out in the optical bands. They primarily used the $y$-band from PS1 and $z$-band from DELS as dropout bands. They also utilized infrared bands from UKIDSS/UHS/VHS and \textit{WISE} as detection bands. They further used the infrared colors to remove the contamination from galactic brown dwarfs. Additionally, they used forced aperture photometry and visual inspection to reject contaminants and bad photometry measurements. We observed 83 candidates from these searches in our Keck programs, but none of them were classified as quasars.

\subsubsection{PS1COLOR: PS1 \& VIKING}

Ba\~nados et al. (in prep) is the third publication reporting the discoveries of $z\sim 6$ quasars in the PS1 survey. They also includes results from a search in the VIKING survey, with an area of 1350 $\rm deg^2$. Both searches are based on color cuts. The procedures of the PS1 search are described in detail in \cite{Banados2016}, and the procedures of the VIKING search are in \cite{Venemans2015}. Ten candidates were observed in our Keck programs, and 2 of them were confirmed as quasars.

\subsubsection{LOFAR: DELS+WISE+LoTSS}

In contrast to the previous quasar searches that only used optical/IR imaging, \cite{Gloudemans2022} used radio detection from the LOFAR Two-metre Sky Survey \citep[LoTSS, 144MHz;][]{Shimwell2017} as the primary selection criterion. The initial sample was built from DELS based on the photometric redshift estimates in \cite{Duncan2022} and optical (DELS-$grz$) and NIR (\textit{WISE}-$W1$) colors. They further required the sources in the initial sample to have a radio detection from LoTSS. The extra constraint on radio emission can remove a large fraction of stellar contaminants. Additionally, they performed SED fitting with both galaxy and quasar templates to narrow down the candidates with inferred redshift. A final sample of 142 candidates were selected for spectroscopic confirmation. Four of them were observed with Keck/LRIS in our programs, and 2 of them were confirmed as quasars.

\subsection{Optical and Near-Infrared Spectroscopy}\label{sec:spectroscopy}

In the two Keck programs, we utilize three different spectrometers to search for high-$z$ quasars, namely LRIS, MOSFIRE and NIRES. Below we provide relevant information for the observation with each instrument.

LRIS is an optical spectrometer with a double spectrometer. For the purpose of high-$z$ quasar confirmation we only use the red part with a pixel scale of	0\farcs123 per pixel. Owing to the recently upgraded red detector \citep{2010SPIE.7735E..0RR} and its working wavelength (optical), LRIS has significantly higher sensitivity for quasar confirmation than the other two spectrometers at the same wavelength. Specifically, we used the 1\arcsec slit, the gold coated 600/10000 grating, and the D680 dichroic in our observation. These result in a wavelength coverage from $\sim 7500\angstrom$ to $\sim 10600\angstrom$ and a spectral resolution of $214\ {\rm km/s}$ at $9700\angstrom$. For each candidate observed with LRIS, we generally take $2\times 300$s exposures to achieve ${\rm S/N}\sim3$ per resolution element. Moreover, we did not dither between exposures during the LRIS observations. In ideal conditions, the upper bound of the wavelength coverage enables us to confirm quasars up to $z\sim 7.6$. But in order to test whether we can reach such a high redshift in practice, we perform simulations on the observations with LRIS, as described in Section \ref{sec:refine}. 

MOSFIRE is an infrared multi-object spectrometer. While it can detect quasars with higher redshift than LRIS, it also suffers from the higher infrared sky background. For MOSFIRE observations, we dithered along the slit to enable data reduction via image differencing as is customary in the NIR. We take MOSFIRE spectroscopy in the $Y$-band with a slit width of 1\arcsec. The wavelength coverage of the $Y$-band is from $9716\angstrom$ to $11250\angstrom$. The spectral resolution with this setting is $\sim 5\angstrom$. The exposure time for each frame is 150s and each dithering sequence is usually consist of 4 frames in ABAB pattern. We take at least one sequence for each target, thus the minimal total exposure time is 600s. 

NIRES is an infrared echellette spectrometer, whereas the previous two are both multi-object spectrometers. The wavelength coverage of NIRES is $0.8-2.4\micron$ in echelle format, spanning five orders. The slit width of NIRES is fixed to 0\farcs55, leading to a resolution of $120\ {\rm km/s}$ at $1\micron$ in the bluest orders. In NIRES observation, we dither along the slit following an ABBA pattern with 
an exposure time of 300s or 360s per frame in the sequence. The typical total exposure time for a target is thus 1200s.  

\begin{figure}
    \centering
    \includegraphics[width=\columnwidth]{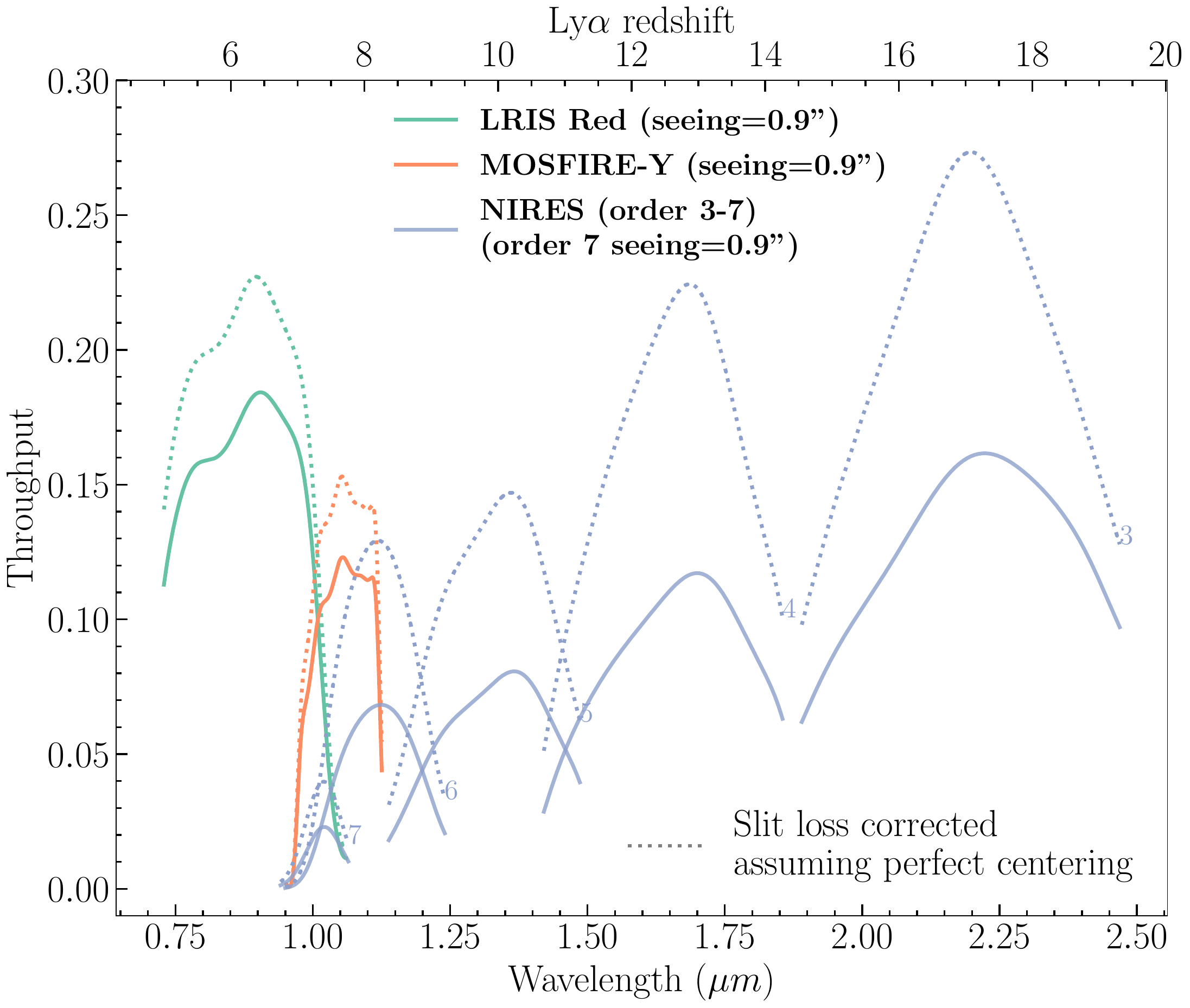}
    \caption{The solid lines are throughput curves of the three spectrometers, corresponding to an 0\farcs9 seeing for LRIS, MOSFIRE, and order 7 of NIRES. The throughput of LRIS was originally measured with an 1\arcsec slit width in LRIS-2203 run, with a seeing of 0\farcs74. The MOSFIRE Y-band throughput was obtained with a 1\arcsec slit width and a 0\farcs76 seeing in MOSFIRE-2201 run. The NIRES throughput here was measured with a 0\farcs55 slit width in NIRES-1905 run with a seeing of 0\farcs9 in the order 7. The LRIS and MOSFIRE throughput curves are then scaled to a seeing of 0\farcs9. The dotted lines are throughput curves corrected for the slit loss, assuming a perfect centering of the star on the slit. The upper axis shows the corresponding Ly$\alpha$ ($1216$ $\angstrom$) redshift of high-$z$ quasars that they can reach. For all instruments the throughput curves are determined from observation of the standard star GD153.}
    \label{fig:throughput}
\end{figure}

In Figure \ref{fig:throughput} we show the throughput curves of LRIS, MOSFIRE, and NIRES 1) under a condition of 0\farcs9 seeing, and 2) corrected for the slit loss, assuming the standard stars are perfectly centered in the slit. The throughput of LRIS was originally measured with an 1\arcsec slit width in LRIS-2203 run, with a seeing of 0\farcs74. The MOSFIRE Y-band throughput was obtained with a 1\arcsec slit width and a 0\farcs76 seeing in MOSFIRE-2201 run. The NIRES throughput here was measured with a 0\farcs55 slit width in NIRES-1905 run with a seeing of 0\farcs9 in the bluest order. The throughput curves are then scaled to 0\farcs9 or corrected for the slit loss following the procedures described in Section \ref{sec:reduction}.

\subsection{Spectroscopic Data Reduction with \texttt{PypeIt}}\label{sec:reduction}

We use the Python Spectroscopic Data Reduction Pipeline \cite[\texttt{PypeIt};][]{pypeit:joss_pub} to reduce all the data in a consistent manner. In the following we will briefly describe the reduction procedures using \texttt{PypeIt}. A more comprehensive version can be found in the \texttt{PypeIt} documentation\footnote{\href{https://pypeit.readthedocs.io/en/release/}{https://pypeit.readthedocs.io/en/release/}}. 

First, we perform a standard automated \texttt{PypeIt} reduction with the script \texttt{run\_pypeit} on all the science frames. This includes the following steps:

\begin{enumerate}
    \item Basic image processing including gain correction, bias subtraction, dark subtraction, and flat fielding.
    \item Constrction of the wavelength solutions and the wavelength tilt models based on either arc (mostly for LRIS) or science frames (i.e. using sky OH lines, mostly for NIR instruments).
    \item Cosmic rays removal with the L. A. COSMIC algorithm\citep{2001PASP..113.1420V}.
    \item Sky subtraction, which is split into multiple steps: The first global sky subtraction is performed with a B-spline fitting procedure \citep[e.g.][]{2003PASP..115..688K}, following by the first object detection. Then the second global sky subtraction is performed, masking the objects previously detected. The second object detection is then implemented. Finally, a local sky-subtraction is applied on all the objects.
\end{enumerate}

After the general reduction, we calculate the sensitivity function using the observation on the standard star with \texttt{PypeIt} script \texttt{pypeit\_sensfunc}. With the sensitivity function, we perform the following procedures to produce fully reduced spectra:

\begin{enumerate}
    \item Flux calibration that converts the spectra in pixel counts into flux with the sensitivity function using \texttt{PypeIt} script \texttt{pypeit\_flux\_calib}.
    \item Co-addition of the multiple frames of the same targets together to achieve higher S/N. For bright objects, we combine the 1d spectra directly using \texttt{PypeIt} script \texttt{pypeit\_coadd\_1d}. For faint objects, we combine their 2d spectra following the dithering pattern with \texttt{PypeIt} script \texttt{pypeit\_coadd\_2d} and extract the 1d spectra afterward. All the 1d spectra are extracted with boxcar and optimal extraction method \citep{1986PASP...98..609H}.
    \item Telluric correction with \texttt{PypeIt} script \texttt{pypeit\_tellfit}. For confirmed quasars, we correct for telluric absorption with quasar model. For the other candidates, we use a polynomial model with an order of 7.
\end{enumerate}

The throughput curves in Figure \ref{fig:throughput} are intermediate products of the reduction. To derive the throughput curves with full transmission or a given seeing, we start with assuming that the standard stars are at the center of the slits, and then use the fitted full width half maximum (FWHM) profiles\footnote{The fitting is done by \texttt{PypeIt} automatically.} of the targets and the slit width to correct for the slit losses. To scale to another seeing, we use the ratio between the desired seeing and the mean FWHM of the trace to scale the FWHM profile and repeat the previous calculation again.

All the \texttt{PypeIt} files to reproduce the reduction are publicly available in the Github repository given at the end of the paper.

\subsection{Classification of the Candidates}\label{sec:classification}

After obtaining the fluxed and telluric-correct spectrum for each candidate, the next step is to classify them. We first visually select out quasars and inconclusive ones, i.e. potential quasars but need re-observation to confirm. Then, we use stellar templates (late M types to T type) to fit the spectra of all the other targets. We visually inspect the fitting and identify the sources with a good fit as brown dwarfs. The others are then non-quasar objects with unknown classification. The detailed classification scheme and criteria are as follows: 

\begin{enumerate}
    \item QSO: A sharp continuum break which in most cases is accompanied by an emission line can be identified in the spectrum. We classify this kind of objects as high-$z$ quasar, and the line as Lyman-$\alpha$ emission line.
    \item INCONCLUSIVE: A drop in the flux can been identified in the spectrum, however, the break is too close to the blue end of the spectrum (thus all of them are from NIR observation), and the spectrum is fainter than expected based on the corresponding $m_J$ (see Section \ref{sec:refine}). We classify objects with these characteristics as inconclusive targets, which require re-observation with LRIS to confirm their true identities.
    \item STAR: To select stellar objects from the sources that do not belong to QSO or INCONCLUSIVE class, we fit the spectra of them with existing brown dwarf spectra from \cite{1991ApJS...77..417K} and \cite{1999ApJ...519..802K}, with the spectral type ranging from late M types to early T types.
    We apply standard $\chi^2$ fitting to all the sources that do not belong to the first two types, and determine the best fit spectral type as the one with the minimal $\chi^2$. We visually inspect the best fit result and classify the sources with a good fit as STAR\footnote{We make use of the chi-square per degree of freedom when determining a good fit, but we do not use it as a determination criteria.}. The best fit spectral type is also appended to each STAR type target.
    \item UNQ: We classify the rest of the sources as UNQ, short for unknown non-quasar. Many of UNQ objects also show gradually declining shape towards the blue end or absorption features like brown dwarfs, but the S/N of their spectra are too low to support reliable STAR classification. 
    \item FAIL: Sources in this type do not have corresponding final 1d spectra due to several reasons. The major cause is the absence of the target trace in 2d spectrum, most probably due to very bad observing conditions. Alternatively, the coordinates for these targets were incorrect or the acquisition procedure failed. Also a few sources were close to the moon when they were observed, which might explain some of the failures. 
\end{enumerate}

\section{High-\texorpdfstring{\MakeLowercase{$z$}}{z} quasar candidate archive}\label{sec:result}

In this section, we present the content of the HzQCA and analyze the contaminant population with it. In Section \ref{sec:result_spec}, we display reduced 2d and 1d spectra of all the new quasars (QSO) and INCONCLUSIVE type objects, as well as several spectra of the STAR and UNQ types as examples. In Section \ref{sec:result_cat}, we describe the contents of the HzQCA catalog. In \ref{sec:result_relflux}, we analyze the distribution of our spectroscopically identified brown dwarfs in color space and compare with empirical brown dwarf models that have been adopted in the quasar selection literature. 

\subsection{Reduced Spectra}\label{sec:result_spec}

\subsubsection{New QSOs}\label{sec:qso}

\begin{figure*}
    \centering
    \includegraphics[width=0.9\linewidth]{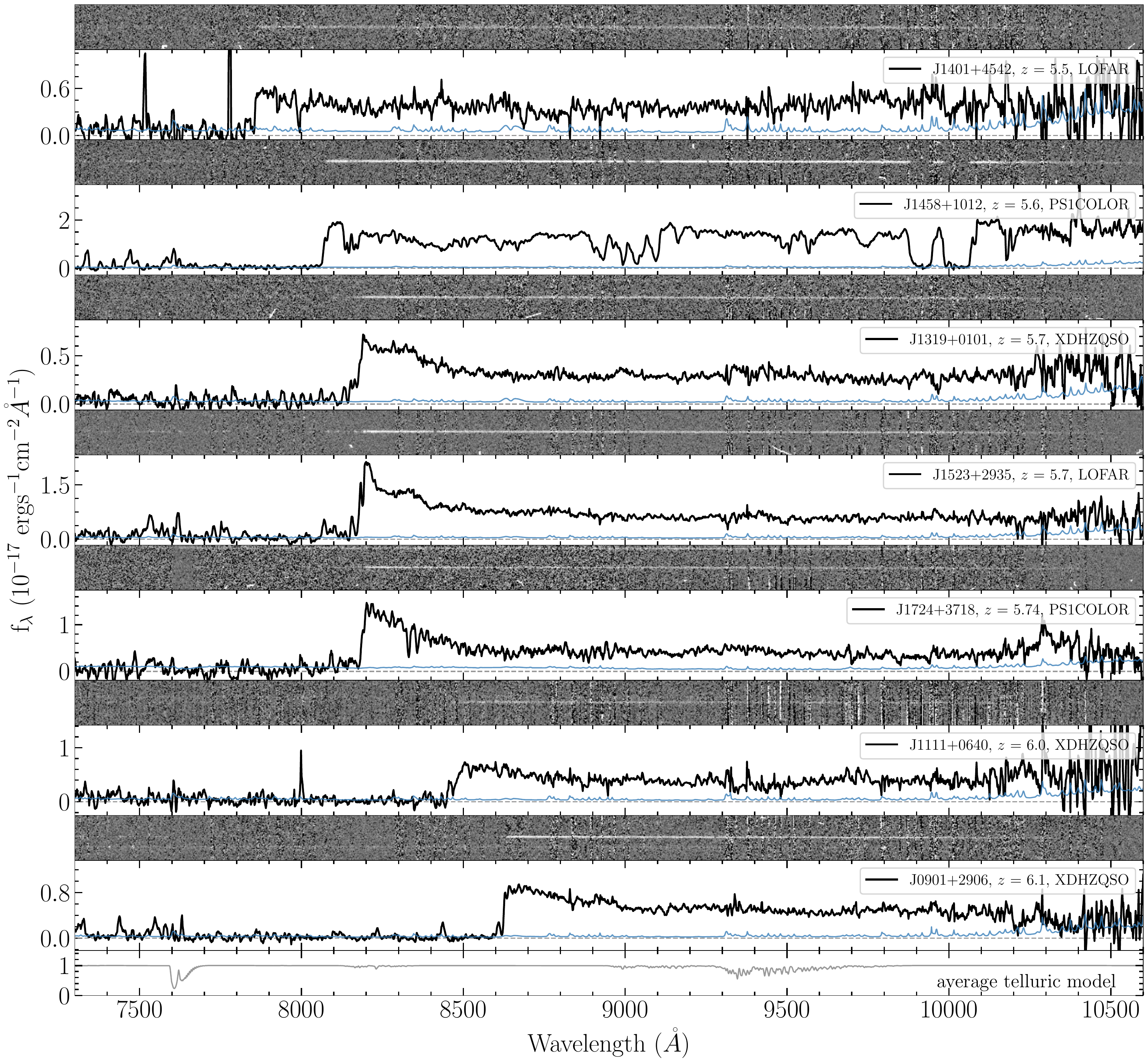}
    \caption{Reduced 2d and 1d spectra of all the new quasars. All of them were observed with LRIS. The reduced 2d spectra are sky-subtracted, but not flux-calibrated and telluric-corrected. The 1d spectra here are flux-calibrated and telluric-corrected. The blue curves are the noise vectors. The grey curve in the bottom panel is the average telluric model. The spectra and noise vectors are smoothed with a 5 pixel boxcar filter, using the inverse variance as relative weights.}
    \label{fig:qso_1d2d}
\end{figure*}

\begin{table*}
	\centering
	\caption{New QSOs discovered in this work. The `Survey' column refers to the surveys mentioned in Section \ref{sec:survey}. References for the acquisition of the photometry measurements are also described in Section \ref{sec:survey}. Here $z$ denotes the $z$-band magnitude, not the redshift.}
	\label{tab:new_qso}
	\begin{tabular}{cccclccccccc}
		\hline
        Name & RA & Dec & Redshift & Survey & $z$ & $Y$ & $J$ & $H$ & $K$ & $W1$ & $W2$ \\	
		\hline
J0901+2906 & 135\degr24\arcmin32\farcs04 & 29\degr 06\arcmin55\farcs44 &      6.10 &     XDHZQSO &  20.86 &  20.51 &  20.59 &  21.43 &  20.05 &   20.11 &   20.01 \\

J1111+0640 & 167\degr51\arcmin17\farcs28 & 06\degr40\arcmin54\farcs12 & 6.00 &     XDHZQSO &  20.83 &  21.46 &  20.75 &  20.48 &  21.20 &   20.66 &   20.40 \\

J1319+0101 & 199\degr51\arcmin52\farcs92 & 01\degr01\arcmin56\farcs28 & 5.70 &     XDHZQSO &  21.26 &  21.55 &  21.43 &  21.09 &  21.04 &   21.66 &   21.93 \\

J1401+4542 & 210\degr20\arcmin12\farcs84 & 45\degr42\arcmin53\farcs28 & 5.50 &     LOFAR &  21.16 &  -- &  -- &  -- &  -- &   -- &   -- \\

J1458+1012 & 224\degr39\arcmin02\farcs52 & 10\degr12\arcmin49\farcs68 & 5.60 &     PS1COLOR &  19.90 &  -- &  -- &  -- &  -- &   -- &   -- \\

J1523+2935 & 230\degr52\arcmin40\farcs08 & 29\degr35\arcmin39\farcs84 & 5.70 &     LOFAR &  20.26 &  -- &  -- &  -- &  -- &   -- &   -- \\

J1724+3718 & 261\degr07\arcmin28\farcs92 & 37\degr18\arcmin21\farcs96 & 5.74 &     PS1COLOR &  21.12 &  -- &  -- &  -- &  -- &   -- &   -- \\
		\hline
	\end{tabular}
\end{table*}

In this work, we discover 7 new $z>5.5$ quasars, which are selected by the XDHZQSO, PS1COLOR and LOFAR searches. We display their reduced 2d and 1d spectra in Figure \ref{fig:qso_1d2d}, and list them in Table \ref{tab:new_qso}. The reduced 2d spectra are sky-subtracted, but not flux-calibrated or telluric-corrected. The 1d spectra here are flux-calibrated and telluric-corrected. They are also smoothed with a 5 pixel boxcar filter, but using the inverse variance as relative weights. The new quasars from PS1COLOR and LOFAR searches are reported in Ba\~nados et al. (in prep) and \cite{Gloudemans2022} separately. The redshift reported here have been determined by visual inspection.


\subsubsection{STAR, UNQ, and INCONCLUSIVE}

\begin{figure*}
    \centering
    \includegraphics[width=0.9\linewidth]{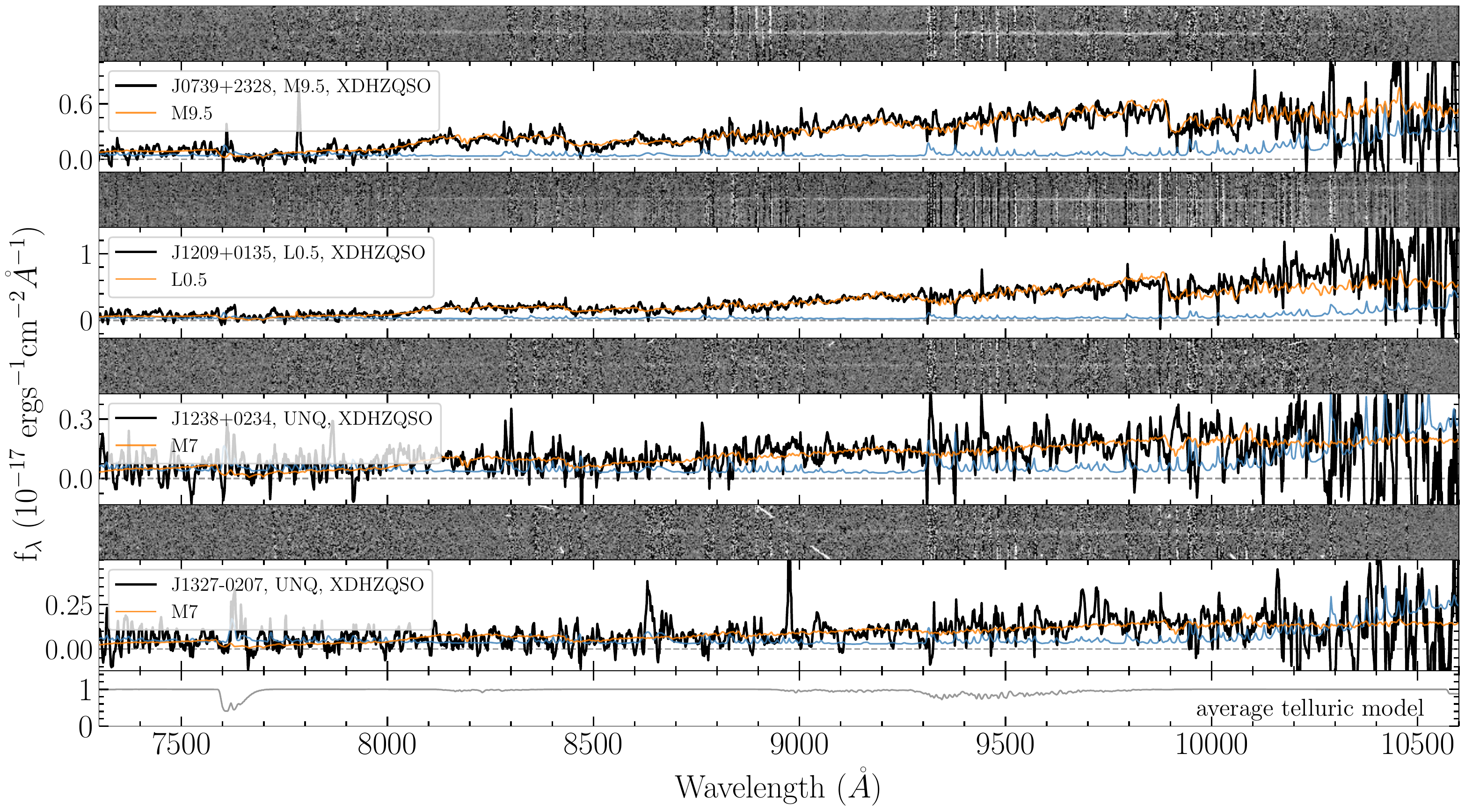}
    \caption{Reduced 2d and 1d spectra of several STAR and UNQ types objects observed with LRIS. The reduced 2d spectra are sky-subtracted, but not flux-calibrated or telluric-corrected. The 1d spectra here are flux-calibrated and telluric-corrected. The blue curves are the noise vectors, and the orange curves are the brown dwarf spectra with the lowest reduced $\chi^2$ values. The grey curve in the bottom panel is the average telluric model. The spectra and noise vectors are smoothed using the inverse variance with a smoothing window of 5.}
    \label{fig:noqso_1d2d}
\end{figure*}

\begin{figure*}
    \hspace{0.25cm}
    \includegraphics[width=0.925\linewidth]{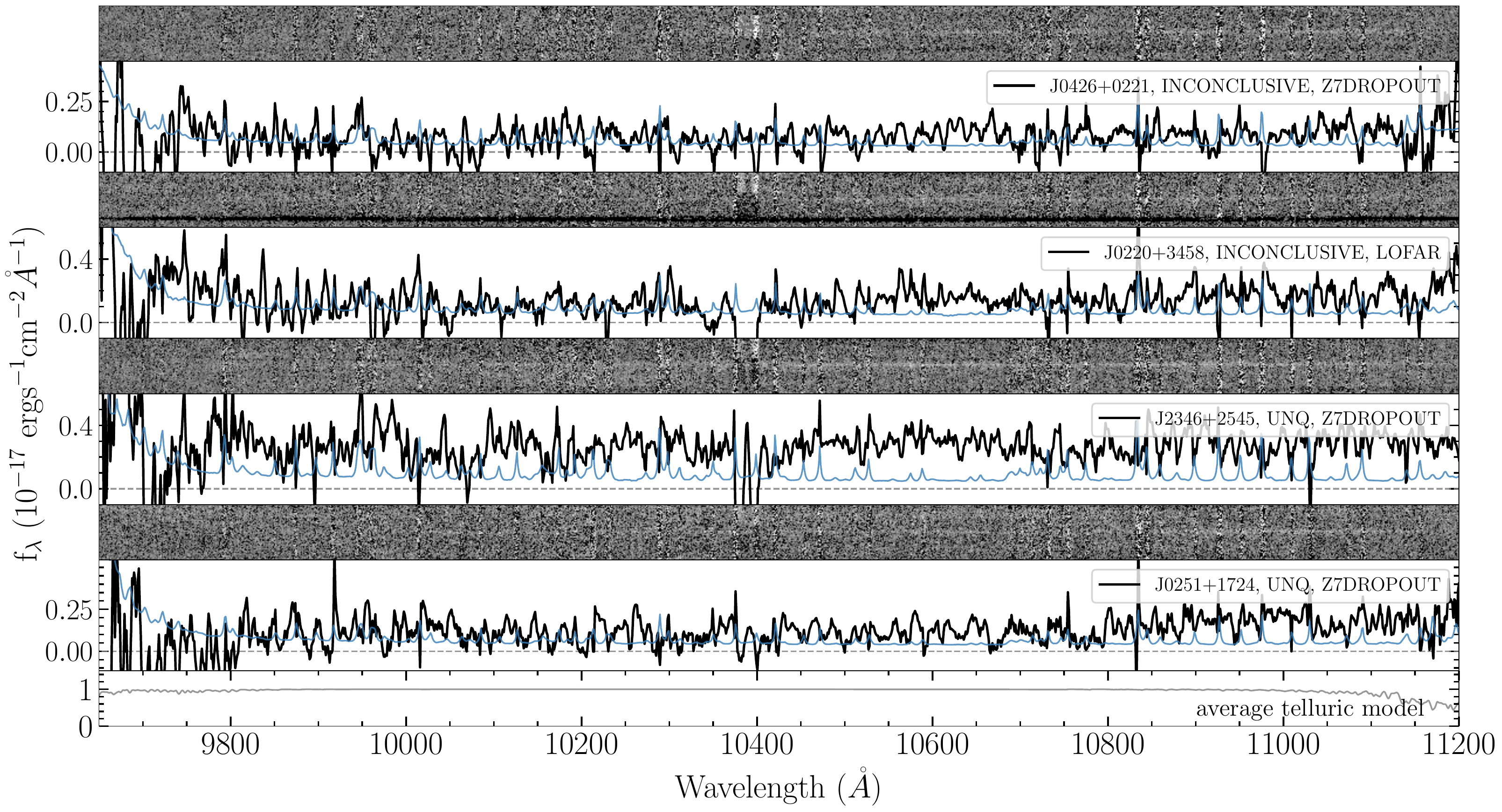}
    \caption{Reduced 2d and 1d spectra of two INCONCLUSIVE type and two UNQ type objects observed with MOSFIRE. The reduced 2d spectra are sky-subtracted, but not flux-calibrated or telluric-corrected. The 1d spectra here are flux-calibrated and telluric-corrected. The blue curves are the noise vectors. The grey curve in the bottom panel is the average telluric model. The spectra and noise vectors are smoothed using the inverse variance with a smoothing window of 5.}
    \label{fig:mosfire_1d2d}
\end{figure*}

\begin{figure*}
    \centering
    \includegraphics[width=0.9\linewidth]{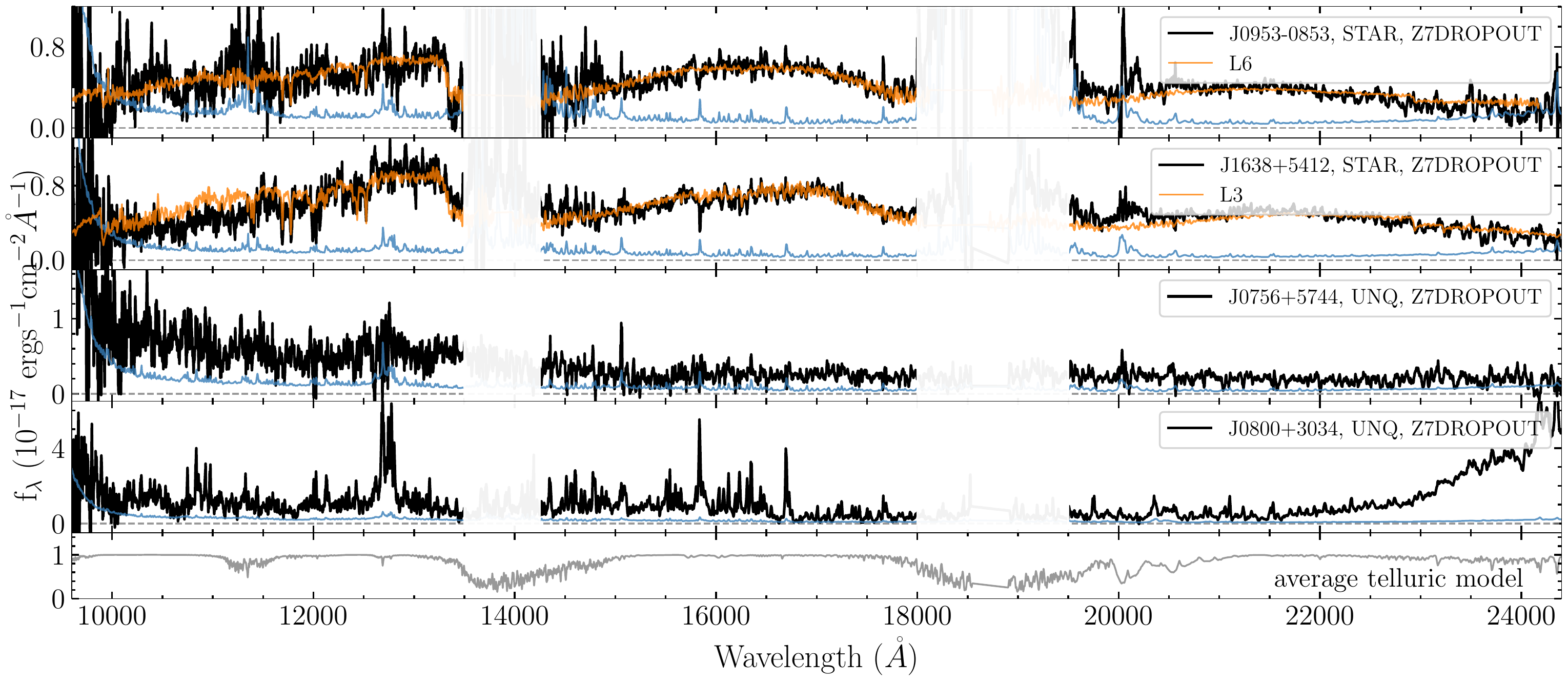}
    \caption{Reduced 2d and 1d spectra of two STAR type and two UNQ objects observed with NIRES. The 1d spectra here are flux-calibrated and telluric-corrected, and the orange curves are the brown dwarf spectra with the lowest reduced $\chi^2$ values. The blue curves are the noise vectors. The grey curve in the bottom panel is the average telluric model. The spectra and noise vectors are smoothed using the inverse variance with a smoothing window of 9. We do not show the 2d spectra here because NIRES has multiple orders.}
    \label{fig:nires_1d2d}
\end{figure*}

In Figure \ref{fig:noqso_1d2d}, we show four examples of STAR and UNQ types objects observed with LRIS. The best-fit brown dwarf templates are also shown (orange lines). The first two objects that have a good fitting results are labeled as STAR and the other two are UNQ. Like many other UNQ type objects, J1238+0234 in this figure shows gradually declining continuum towards the blue end, similar with a brown dwarf. But the S/N of these spectra are not high enough to support a reliable brown dwarf classification.

In Figure \ref{fig:mosfire_1d2d}, we show the spectra of two INCONCLUSIVE type and two UNQ type objects observed with MOFIRE. The spectra of the two INCONCLUSIVE type objects drop near the blue end. However, as the throughput curve of MOSFIRE drops steeply $\lesssim 1\micron$ as well, we cannot confidently identify such drops in fluxes as continuum breaks. It is also possible that some of the UNQ type objects observed with MOSFIRE are actually quasars with lower redshift, or with a reddened spectrum. But we will need higher S/N optical spectra to make such classification.

Finally, two STAR type and two UNQ type objects observed with NIRES are displayed in Figure \ref{fig:nires_1d2d}. We can see that the STAR type objects are cooler brown dwarfs than those in LRIS runs (see Figure \ref{fig:noqso_1d2d}). This is expected as all the candidates in NIRES observation runs are from the Z7DROPOUT searches (see Section \ref{sec:Z7DROPOUT}, where most of the contaminants are late L type and T type brown dwarfs. We note that the flux in the red end of J0800+3034 is not real, but a saturating issue.

The spectra of other sources that belong to STAR and UNQ types (there are only two INCONCLUSIVE type objects, and are all displayed in Figure \ref{fig:mosfire_1d2d}) are provided in Appendix \ref{sec:spectrum}.

\subsection{The Catalog}\label{sec:result_cat}

In this section we describe the contents of the HzQCA. The classification results, following Section \ref{sec:classification}, are summarized in Table \ref{tab:summary}. There are 7 new quasars discovered in these observation and 3 of them are newly reported in this paper (see Section \ref{tab:new_qso}). We confirm that no quasar was missed during the `quick-look' classification on the fly. There are 51 sources classified as brown dwarfs, while 74 sources are UNQ objects. Two candidates in MOSFIRE runs fall into the INCONCLUSIVE class. Finally, 42 objects are labeled as FAIL. Most of the objects in the FAIL type were observed in the first two runs with MOSFIRE, when the observing strategy was under testing. The observing condition of the MOSFIRE-1911 run was also not good, leading to more objects with lower S/N.

\begin{table}
	\centering
	\caption{Summary of the classification.}
	\label{tab:summary}
	\begin{tabular}{lcr}
		\hline
		Type & Number & Fraction \\
		\hline
		QSO & 7 & $4\%$ \\
		INCONCLUSIVE & 2 & $1\%$ \\
		STAR & 51 & $29\%$ \\
		uNQ & 72 & $41\%$ \\
		FAIL & 42 & $24\%$ \\
		\hline
		Total & 174 & 100\% \\
		\hline
	\end{tabular}
\end{table}

The format of the full catalog is presented in Table \ref{tab:result}. The full archive is provided in the electronic version of the journal. 

\begin{table*}
	\centering
	\caption{Part of the HzQCA. The labels are assigned following the procedures in Section \ref{sec:classification}. The surveys are those described in Section \ref{sec:survey}. The $J$-band magnitudes are from UKIDSS or VIKING, if available.}
	\label{tab:result}
	\begin{tabular}{ccrccccc}
		\hline
        Candidate & RA & Dec & $m_J$ & Label & Spectral type & Redshift & Survey \\		
        \hline
		\multicolumn{8}{c}{LRIS-2203 (Part)}\\
		\hline
J0810+2352 & 122\degr40\arcmin04\farcm44 & 23\degr52\arcmin35\farcm04 & 20.20 &         STAR &            M9 &       -- &           XDHZQSO  \\
J0850+0146 & 132\degr32\arcmin06\farcm72 & 01\degr46\arcmin42\farcm96 & 20.95 &         STAR &            L2 &       -- &           XDHZQSO  \\
J0901+2906 & 135\degr24\arcmin32\farcm04 & 29\degr06\arcmin55\farcm44 & 20.59 &          QSO &           -- &      6.10 &           XDHZQSO  \\
J0911+0022 & 137\degr55\arcmin04\farcm44 & 00\degr22\arcmin42\farcm60 & 21.29 &         STAR &            M9 &       -- &           XDHZQSO  \\
J0947+0111 & 146\degr58\arcmin52\farcm32 & 01\degr11\arcmin26\farcm88 & 20.57 &         STAR &            M8 &       -- &           XDHZQSO  \\
J1100+0203 & 165\degr00\arcmin43\farcm56 & 02\degr03\arcmin00\farcm36 & 21.52 &          UNQ &           -- &       -- &           XDHZQSO  \\
J1143-0248 & 175\degr59\arcmin01\farcm68 & -02\degr48\arcmin29\farcm52 & 21.76 &          UNQ &           -- &       -- &           XDHZQSO  \\
J1153-2239 & 178\degr20\arcmin23\farcm28 & -22\degr39\arcmin01\farcm08 & -- &         STAR &            M9 &       -- &           PS1COLOR  \\
J1200+0112 & 180\degr00\arcmin47\farcm52 & 01\degr12\arcmin14\farcm40 & 21.67 &          UNQ &           -- &       -- &           XDHZQSO  \\
		\hline
	\end{tabular}
\end{table*}

This archive mainly serves two purposes: firstly, future quasar searches will greatly benefit from the archive as a reference for already observed targets; secondly, we provide a large sample of the contaminant population for systematic analysis. In the following section, we look into the distribution of both quasars and contaminants in the color space.

\subsection{Distributions of the Quasars and Contaminants in Color Space}\label{sec:result_relflux}

It is of significant interest for quasar selection to understand how quasars and their contaminants populate the color space. Armed with the spectroscopically classified targets in our archive, we will now consider their colors with photometry measurements from DELS ($z$), VIKING/UKIDSS ($YJHK_s$/$YJHK$), and un\textit{WISE} ($W1W2$). In the following analysis, we only include sources covered in the VIKING/UKIDSS+DELS+un\textit{WISE} footprints, which are generally the sources from the XDHZQSO search. We use the same photometry data as those used in the XDHZQSO search candidate selection. We also compute relative fluxes besides colors since the observational uncertainties of relative fluxes are closer to Gaussian and because negative fluxes can not be represented with magnitudes \citep[see e.g.][]{Bovy2011,Nanni2022}. We always use $J$-band as the reference band to be consistent with \cite{Nanni2022}, i.e. the relative fluxes are always fluxes in different bands divided by the $J$-band flux.

\begin{figure*}
    \begin{subfigure}{.42\textwidth}
        \centering
        \includegraphics[width=\linewidth]{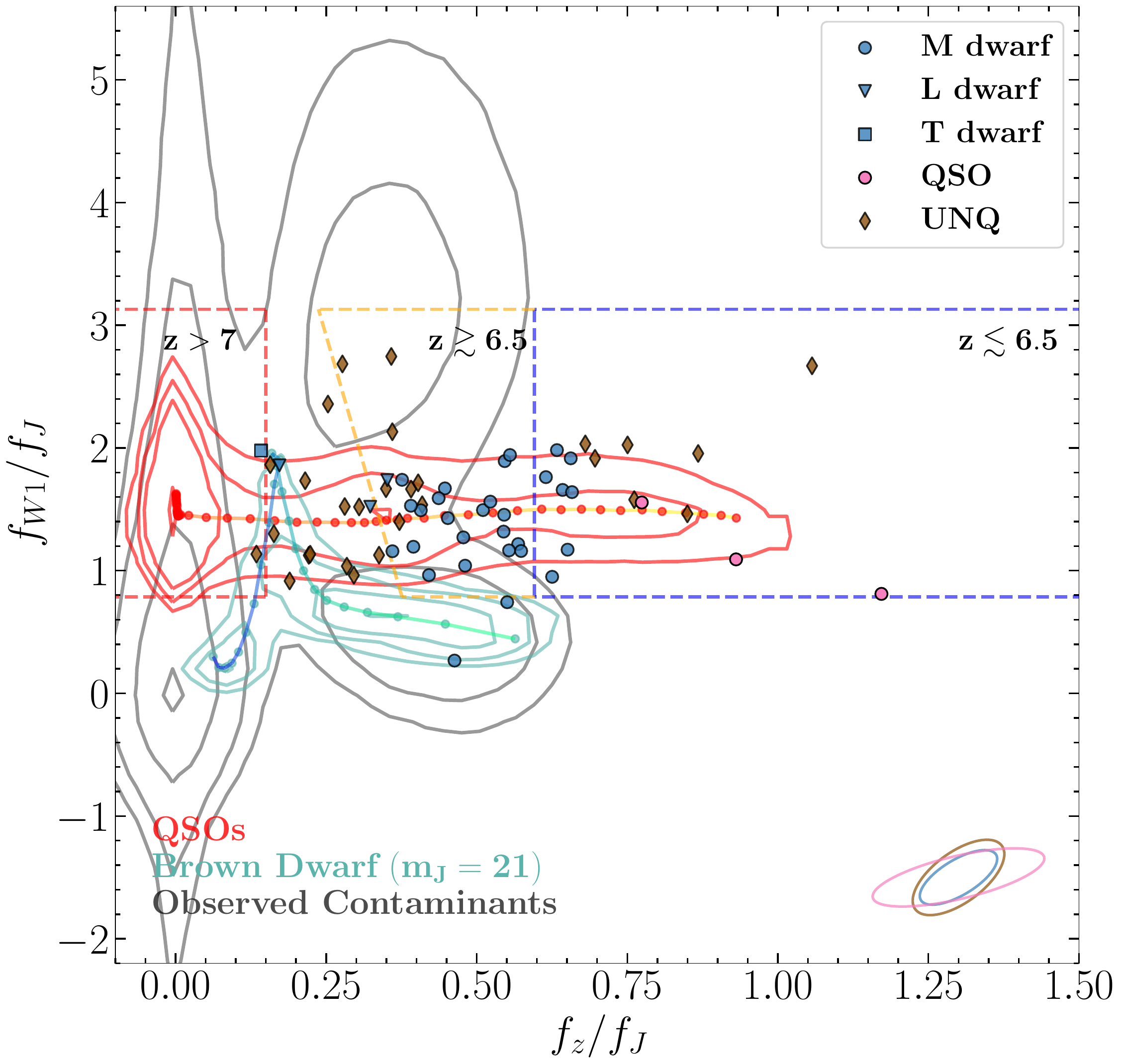}
    \end{subfigure}
    \begin{subfigure}{.41\textwidth}
        \centering
        \includegraphics[width=\linewidth]{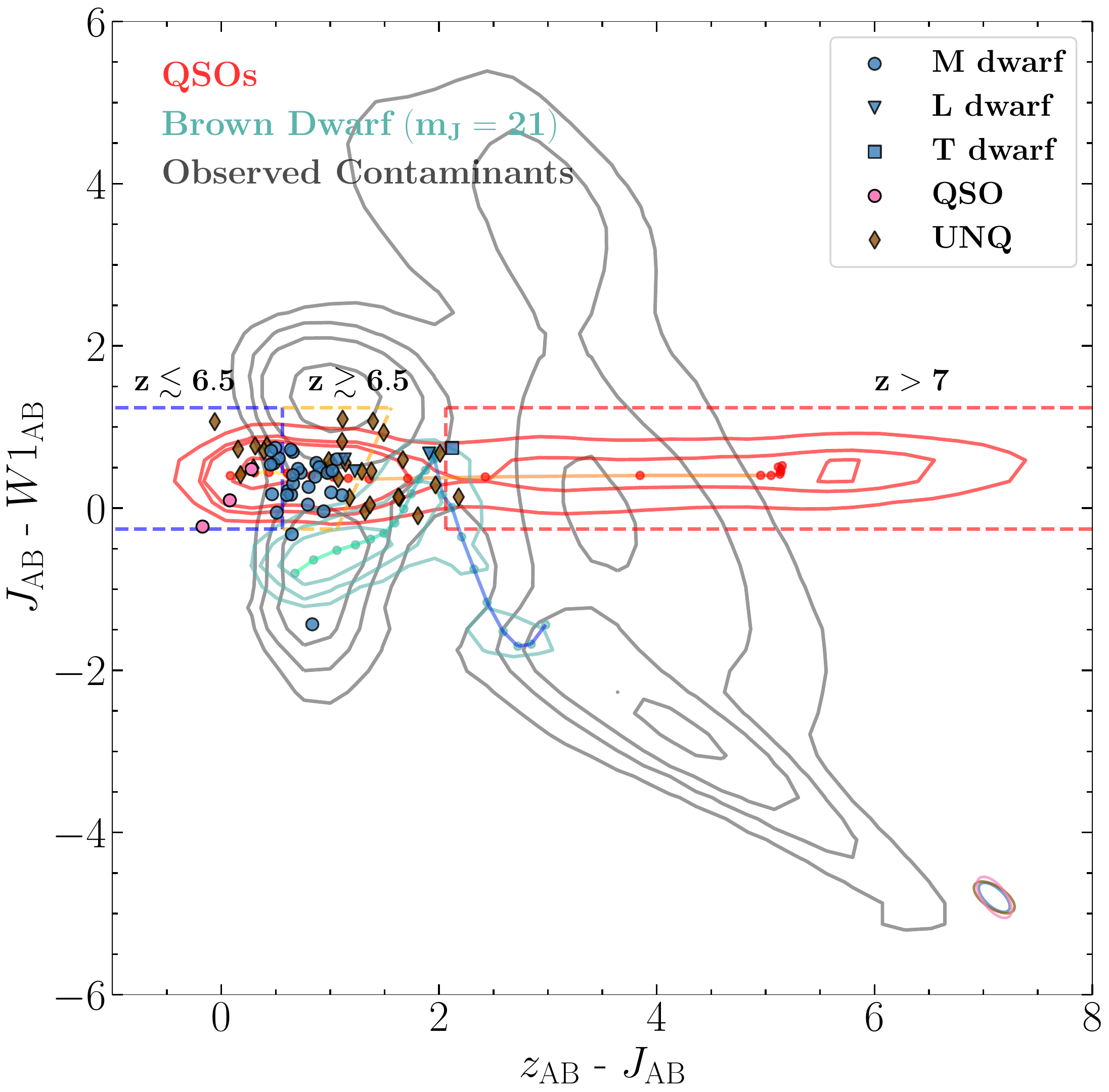}
    \end{subfigure}
    
    \begin{subfigure}{.434\textwidth}
        \hspace{-0.2cm}
        \includegraphics[width=\linewidth]{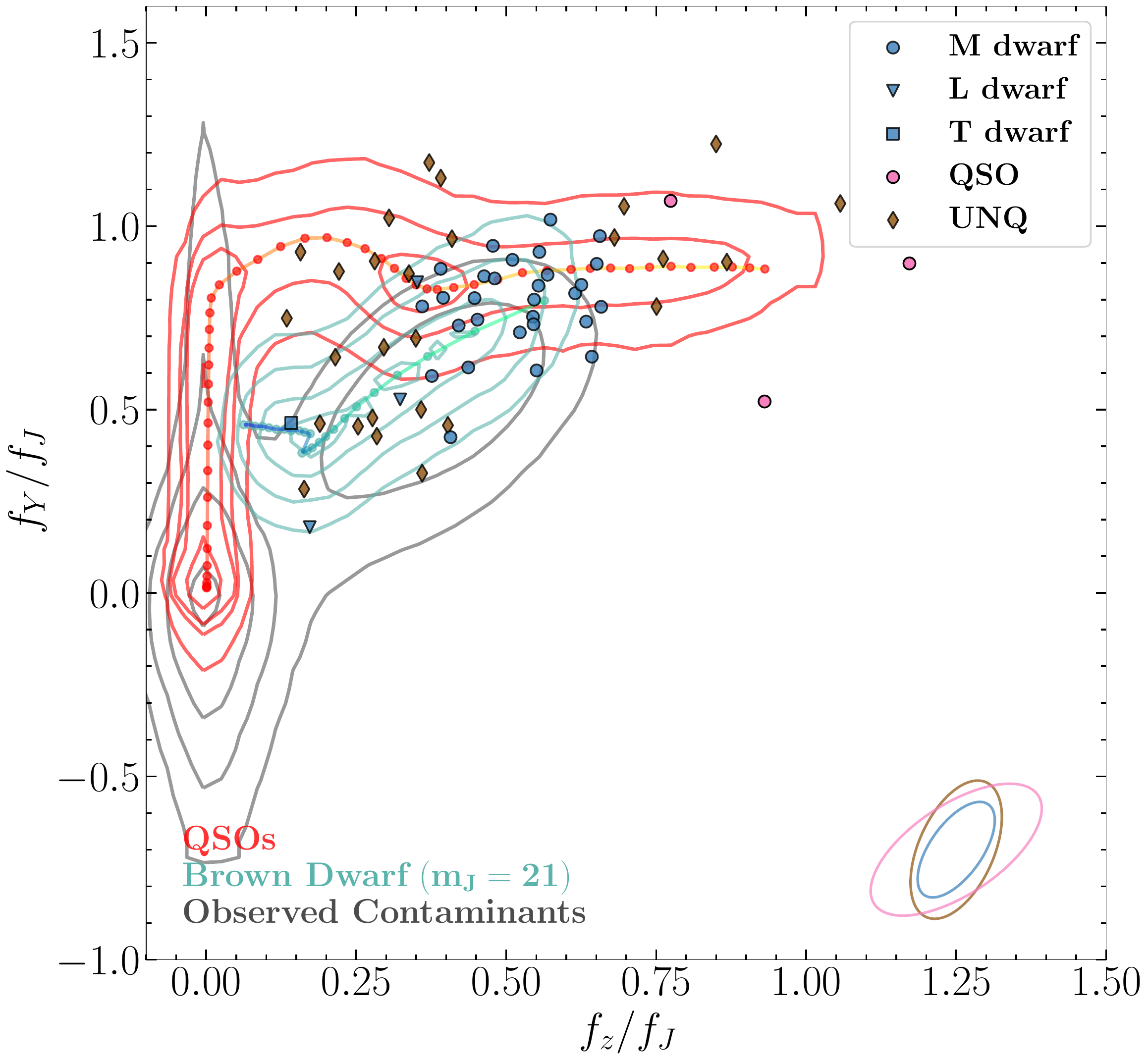}
    \end{subfigure}
    \begin{subfigure}{.41\textwidth}
        \hspace{-0.2cm}
        \includegraphics[width=\linewidth]{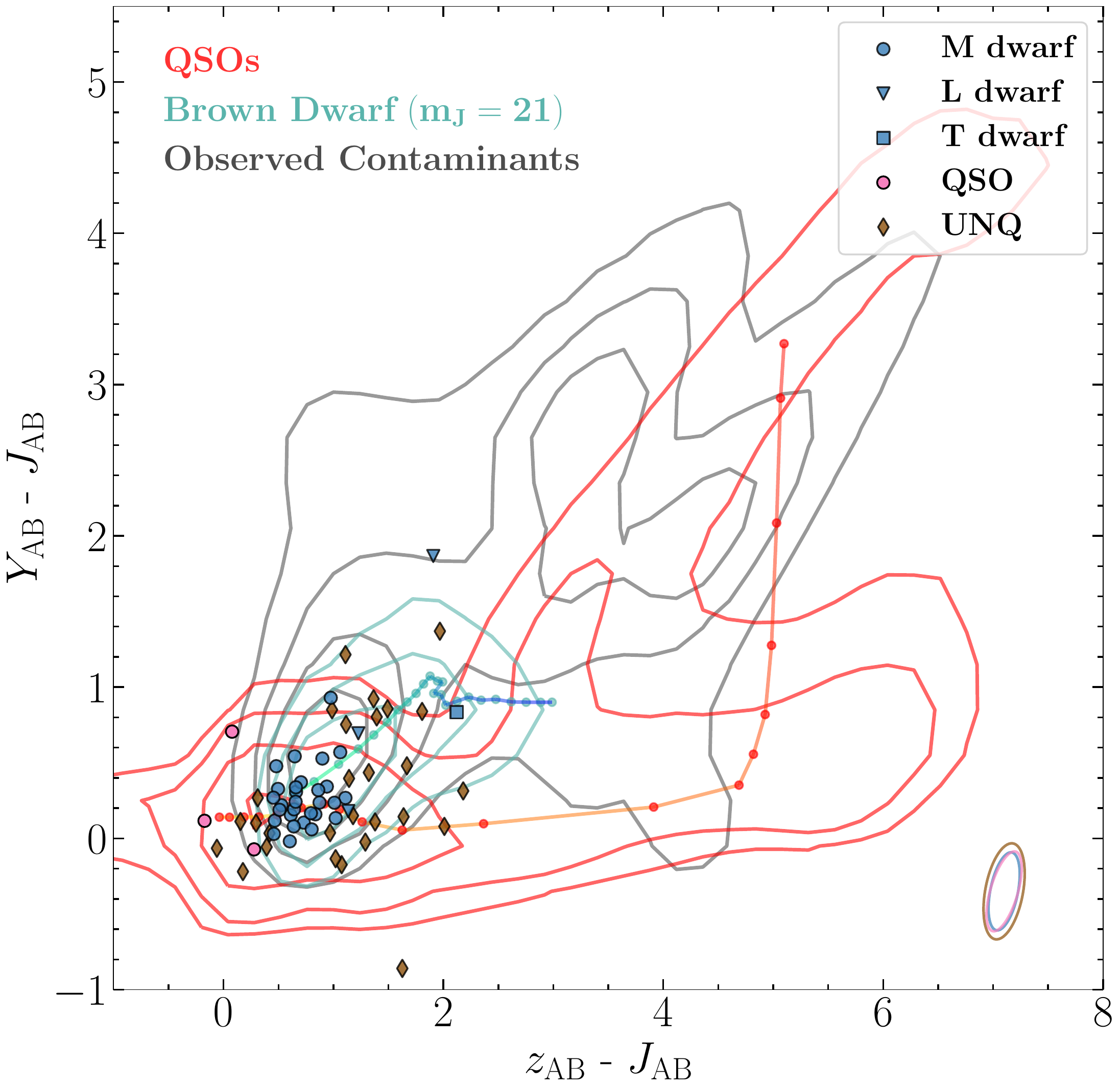}
    \end{subfigure}
    
    \begin{subfigure}{.413\textwidth}
        \hspace{-0.05cm}
        \includegraphics[width=\linewidth]{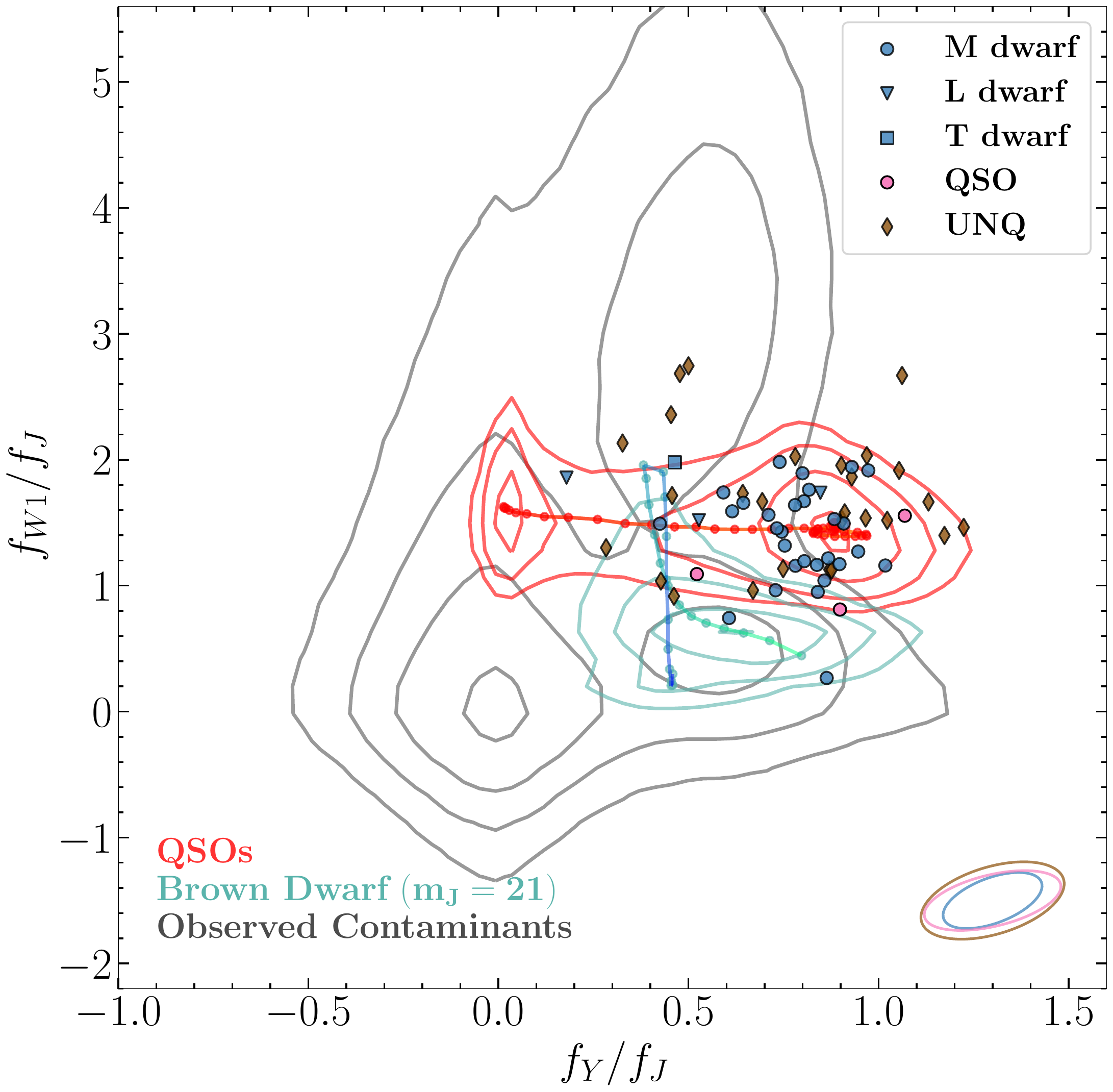}
    \end{subfigure}
    \hspace{0.2cm}
    \begin{subfigure}{.413\textwidth}
        \hspace{-0.1cm}
        \includegraphics[width=\linewidth]{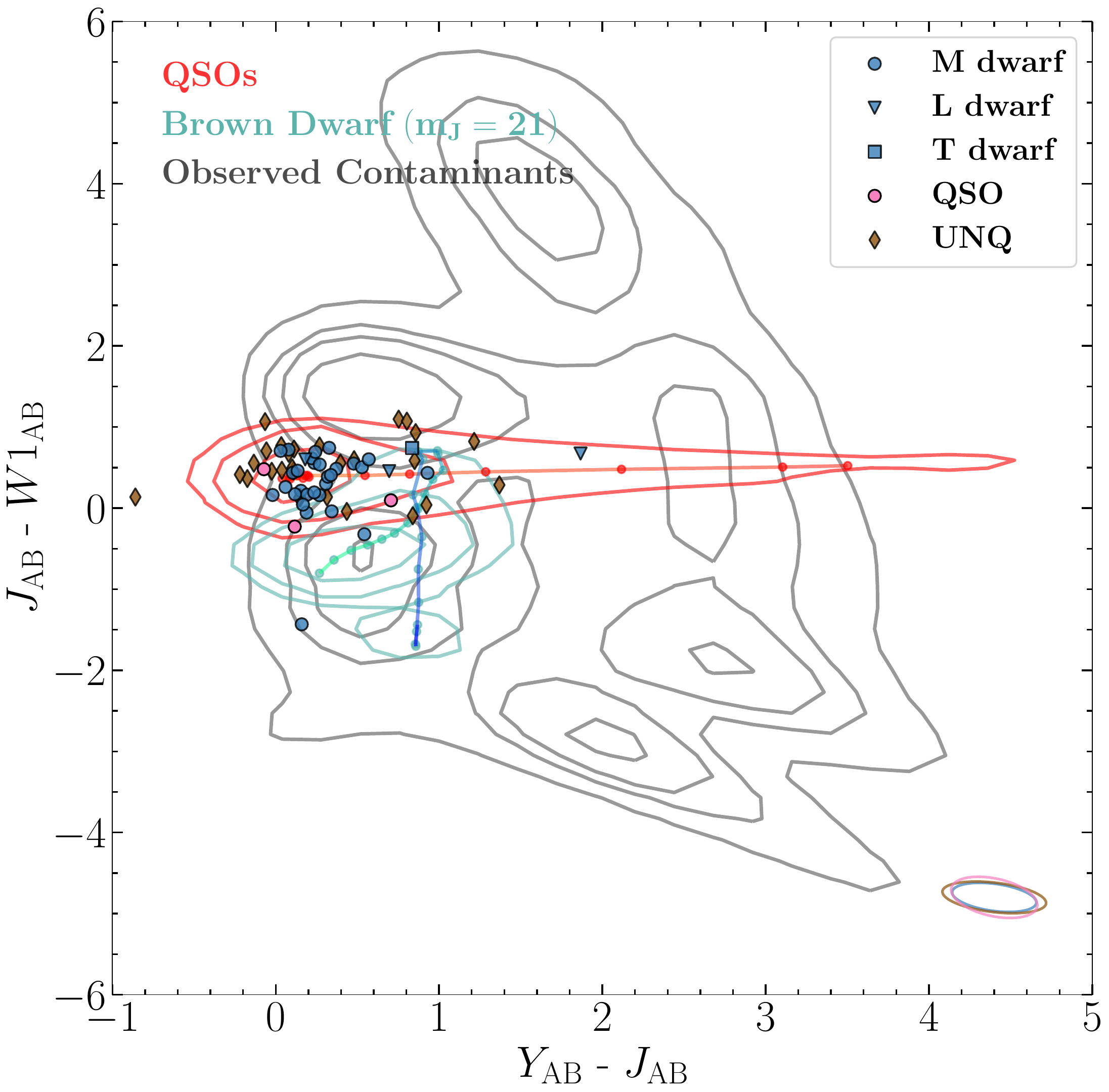}
    \end{subfigure}
    \caption{(Left panels) The relative flux diagrams. The filled symbols are the sources in the HzQCA, including QSO (pink), STAR (blue) and UNQ (brown) types objects. The red contour is the density of 440,000 simulated $6\leq z\leq 8$ quasars generated with \texttt{simqso}. The cyan contour is the density of 100,000 synthetic brown dwarfs, generated from convolving the empirical brown dwarf colors in \protect\cite{Skrzypek2015} and the measurement errors, following the noise modeling procedures described in \protect\cite{Nanni2022}. The cyan line in it indicates different spectral types. The green end of this track is M5 type, while the blue end is the T8 type. The grey contour is the density of the contaminants used to train the contaminant model in XDHZQSO searches (see Section \ref{sec:VIK+UKI}). The dashed lines in the upper panels are the color selection cuts from \protect\cite{Wang2017}. (Right panels) Corresponding color-color diagrams. The levels of the contours are 11.75\%, 39.35\%, 67.53\%, and 86.47\%.} 
    \label{fig:flux_ratio}
\end{figure*}

\begin{figure*}
    \centering
    \includegraphics[width=\linewidth]{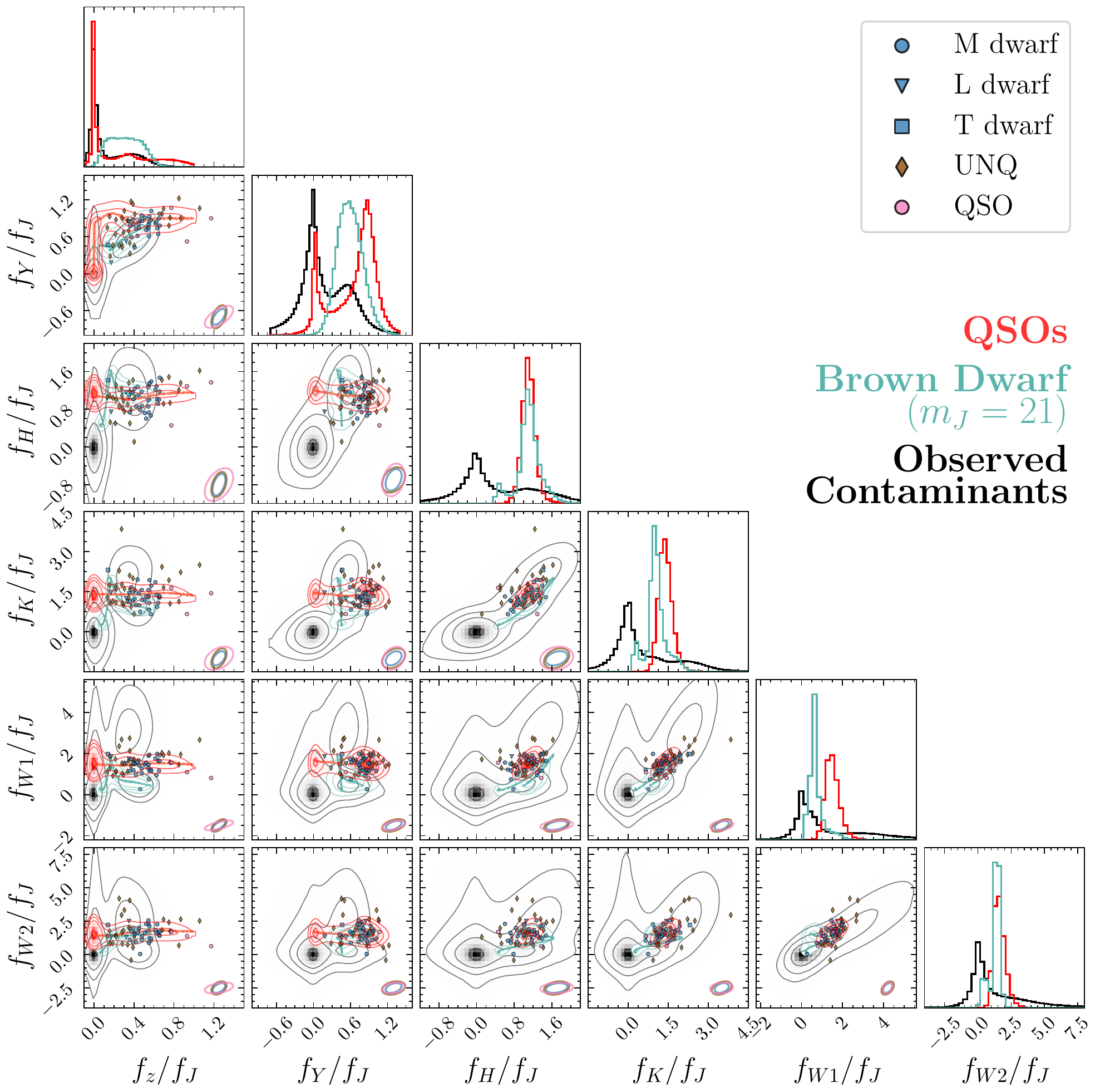}
    \caption{The complete relative flux diagrams. The filled symbols are the sources in the archive, including QSO (pink), STAR (blue) and UNQ (brown) types objects. The red contour is the density of 440,000 simulated $6\leq z\leq 8$ quasars generated with \texttt{simqso}. The cyan contour is the density of 100,000 synthetic brown dwarfs, generated from convolving the empirical brown dwarf colors in \protect\cite{Skrzypek2015} and the measurement errors, following the noise modeling procedures described in \protect\cite{Nanni2022}. The grey contour is the density of the contaminants used to train the contaminant model in XDHZQSO searches (see Section \ref{sec:VIK+UKI}). The levels of the contours are 11.75\%, 39.35\%, 67.53\%, and 86.47\%.}
    \label{fig:flux_ratio_corner}
\end{figure*}

In the left hand side of Figure \ref{fig:flux_ratio}, we show the distributions of the sources in three different relative flux diagrams, i.e. $f_z/f_J$ vs. $f_{W1}/f_J$, $f_z/f_J$ vs. $f_Y/f_J$, and $f_Y/f_J$ vs. $f_{W1}/f_J$. The corresponding color-color diagrams are shown in the right hand side of the figure. The filled symbols are sources belong to the QSO (pink), STAR (blue), UNQ (brown) types. The red contour is the distribution of 440,000 simulated quasars ($6\leq z\leq 8$) generated with \texttt{simqso}. We also convolve the fluxes of all the simulated quasars with measurement uncertainties obtained from real data following the noise modeling procedures described in \cite{Nanni2022}. The cyan contour represents the distribution of brown dwarf flux ratios (colors) based on the empirical brown dwarf colors. \citep{Skrzypek2015} measured the colors of brown dwarfs with different spectral types, from M5 to T8. We then convolve the fluxes (assume $m_J=21$) with errors following \cite{Nanni2022} to get 100,000 synthetic measurements, and built the contour based on these points. This cyan contour is essentially similar to the brown dwarf model adopted in several probabilistic candidate selection methods \citep[e.g.][]{Mortlock2011,Matsuoka2022}. The grey contour is the distribution of all the contaminants used for implementing \texttt{XDHZQSO} in the VIKING/UKIDSS searches (see Section \ref{sec:VIK+UKI}). For a reference, we display the color selection boxes from \cite{Wang2017} in the $f_z/f_J$ vs. $f_{W1}/f_J$ diagram and the corresponding color-color diagram. Finally, we show the average covariance matrices\footnote{The errors are covariant since both relative fluxes (colors) depend on $f_J$ ($m_J$).} of QSO (pink), STAR (blue), and UNQ (brown) types as three ellipses in the lower right position of each panel. The complete relative flux diagrams which includes the $zYHKW1W2$ bands is shown in Figure \ref{fig:flux_ratio_corner}.

From Figure \ref{fig:flux_ratio} we can see that the contaminants in our samples mostly inhabit the quasar locus as is expected, since the \citet{Nanni2022} selection is trained on this quasar locus. Moreover, the spectroscopically confirmed brown dwarfs in our sample nevertheless largely deviate from the general brown dwarf locus indicated by the cyan contours. The nature of these brown dwarfs, which appear to be significant outliers from the brown dwarf locus, is important for quasar selection, since many probabilistic selection methods rely on building empirical models of the brown dwarf in color space \citep[e.g.][]{Mortlock2011,Matsuoka2022}. 

Given our large sample of spectroscopically confirmed brown dwarfs, we are in a good position to test if the brown dwarfs in our sample are consistent with the commonly presumed models (cyan contour). We note an important caveat that our sampling of the brown dwarf is not representative of the general population, as we targeted the sources that were likely to be quasars (red contours). Nevertheless, we can still provide an order of magnitude estimate of whether we expect to find this many outliers with our searches from the brown dwarf locus (cyan contours) given the survey area and depth. The estimate is as follows:
\begin{enumerate}
    \item We first use 1,000,000 synthetic brown dwarfs to fit the density distribution of brown dwarfs in the 6-d relative flux space (see e.g. Figure \ref{fig:flux_ratio_corner}. We used a simple Gaussian Mixture Model (GMM) with 5 Gaussians components. 
    \item With the fitted GMM, we calculate the log probability density (probability in relative flux space) $\ln(p)$ of each synthetic object and obtain a cumulative distribution function of $\ln(p)$. The result shows that the lowest probability out of 1,000,000 samples is $\ln(p)=-44.2$, hence $P(\ln(p)\le -44.2)\approx 0.0001\%$. 
    \item The surface density of brown dwarfs is $\sim 25/{\rm deg}^2$ for $20 < m_J < 22$ \citep{Barnett2019}, and the combined UKIDSS+VIKING survey area is $\sim 5,000{\rm deg}^2$. Therefore, we expect $\sim 125,000$ brown dwarfs in the search area with $20 < m_J < 22$.
    \item The product of the total brown dwarf number with $P(\ln(p)<-44.2)\approx 0.0001\%$ gives us $\sim 0.125$ sources with $\ln(p)<-44.2$ should have been found by our search. However, using the fitted GMM, we estimate $\ln(p)$ of all the brown dwarfs in our archive and find that there are 25 with $\ln(p)<-44.2$. 
\end{enumerate}

According to this rough estimate, it appears that the empirical brown dwarf model (represented by cyan contours in Figure \ref{fig:flux_ratio} and Figure \ref{fig:flux_ratio_corner}) does not represent the entirety of the brown dwarf distribution. This shortcoming is especially clear in the panels that have $K$ and $W1$ measurements, as shown in Figure \ref{fig:flux_ratio_corner}. The origin of this discrepancy, between our spectroscopically confirmed outlier brown dwarfs and the empirical model  is not clear. There are several systematic errors that may contribute to this discrepancy: 1) The errors of these sources are non-Gaussian due to e.g. deblending of overlapped sources, or artifacts in the images, etc, and 2) these sources are variable stellar objects or binary systems; combined with the fact that the photometry data of different surveys were not taken at the same time, the variability might lead to redder colors. For the first issue, visual inspection on the images or repeated photometry are possible approaches, while for the second issue, adopting the empirical model built with all contaminants, or developing a better model of brown dwarf are possible directions.

\section{Improving the observing efficiency with simulated observations}\label{sec:refine}

In the following we assess our observing strategy for future high-$z$ quasar searches with \textit{Euclid} based on the spectroscopic data set we built in this work. Although LRIS has the best performance among the three instruments, the feasibility of using LRIS to observe fainter, more distant quasars that are common in future surveys remains unknown. Moreover, since LRIS is an optical spectrometer (see Figure \ref{fig:throughput}), above what redshift should we shift to IR instruments like MOSFIRE and NIRES, or even \textit{JWST}? These questions arise from our limited experience confirming $z>7$ quasars.

The key to address these issues is to precisely estimate the required exposure time to identify the quasars that we have barely seen, e.g. $m_J\sim 23$, or $z> 7$. To accomplish this goal, we simulate a spectroscopic observation of a quasar, given its redshift and apparent magnitude, and then measure its S/N. We can then use the relation between S/N and exposure time to estimate the required exposure time for identification. We present our simulation procedures in the next section.

\subsection{Synthetic Observations}

We start from a rest-frame quasar template \citep{Selsing2016} and a real 2d and 1d spectrum (stored as \texttt{Spec2DObj} and \texttt{SpecObjs} objects with \texttt{PypeIt}) from the HzQCA. 
The inputs are the redshift ($z$) and apparent magnitude ($m_J$) of the desired quasar. We implement the following steps to simulate the quasar on the 2d spectrum with \texttt{PypeIt} functions and routines:
\begin{enumerate}
    \item We first shift the wavelength of the 1d template by a factor of $(1+z)$, where $z$ is the given redshift. We let the fluxes blueward than the redshifted Ly-$\alpha$ wavelength ($(1+z)\times 1216 \angstrom$) to zero, mimicking the hydrogen absorption of the IGM at high redshifts.
    \item We then measure the $J$-band apparent magnitude $m'_{\rm J}$ of the redshifted quasar template with UKIDSS-$J$ transmission curve. We rescale the spectrum with a factor of $10^{-(m_{\rm J}-m'_{\rm J})/2.5}$ to match with the provided magnitude $m_J$.
    \item After rescaling, we convert the spectrum from flux units into photon counts per second per angstrom with the sensitivity function generated with the standard star observation in the same run as the real 2d spectrum:
    \begin{equation}
        N_\lambda = S_\lambda F_\lambda
    \end{equation}
    where $S_\lambda$ is the sensitivity function. 
    \item Then, we convert $N_\lambda$ into $N_{\rm pix}$ (photon counts per pixel) following:
    \begin{equation}
        N_{\rm pix} = N_\lambda \times \frac{\mathrm{d}\lambda}{\mathrm{d}\rm pix} \times t_{\rm exp}
    \end{equation}
    where $\frac{\mathrm{d}\lambda}{\mathrm{d}\rm pix}$ is the wavelength spacing per pixel which can be obtained from the real 2d spectrum, and $t_{\rm exp}$ is the exposure time of the 2d spectrum.
    
    \item Next, we `duplicate' one of the point sources extracted from the real 2d sky-subtracted frame by shifting its trace\footnote{We refer to the positions of the center of the source spectrum as a function of wavelength (spectral position) as the trace.}
    to a `blank' region of the spectrum free of sources. The variation of FWHM along the trace is estimated by \texttt{PypeIt} for the real source and is also used for the synthetic duplicate source. We then generate a 2d spatial profile along the duplicated trace using the FWHM values, i.e. we put a Gaussian profile centered at each pixel along the trace, and determine the width of the Gaussian with the FWHM value at that pixel.
    \item We interpolate the simulated quasar spectrum onto the wavelength grid corresponding to the duplicated trace in the 2d spectrum. Finally, we scale the Gaussian profiles individually to force the boxcar extracted value of the duplicated trace to be the same as the interpolated spectrum in counts per pixel.
    \item After successfully injecting the simulated quasar into the 2d spectrum, we perform a boxcar extraction to obtain the final 1d spectrum.
\end{enumerate}

The primary benefit of these procedures is that we can use the real empirical noise realization from the 2d spectrum as the noise for our synthetic observations. \texttt{PypeIt} performs both global and local sky-subtraction, as described in Section \ref{sec:reduction}. A fair comparison between real and simulated objects requires performing the local sky-subtraction on the simulated object as well. To do so, we re-ran the reduction on the 2d spectrum, and manually extract the location where we intend to put the simulated quasar as a workaround.

\begin{figure*}
    \centering
    \includegraphics[width=\linewidth]{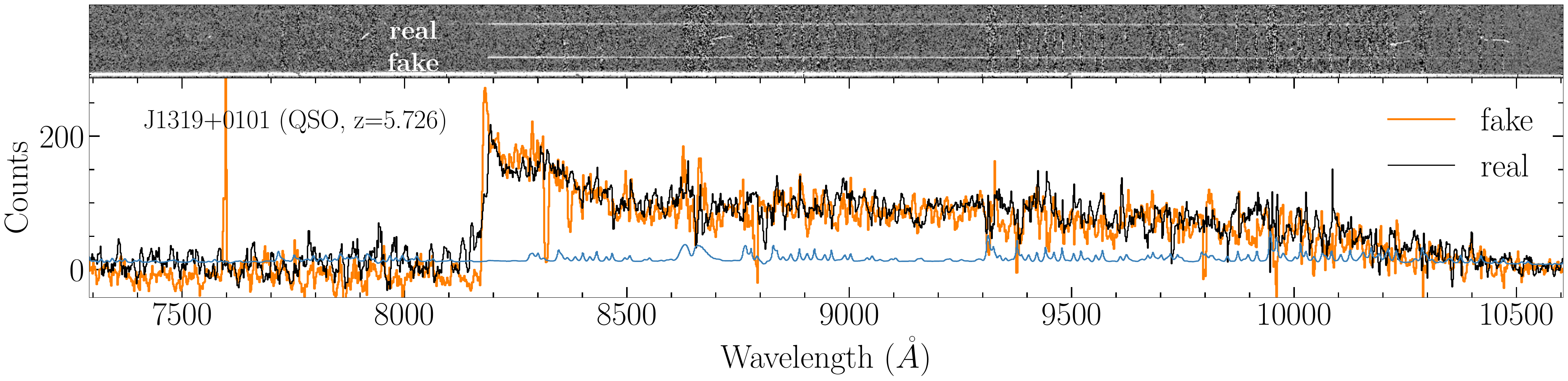}
    \caption{Simulated quasar J1319+0101 in LRIS. (Upper panel) 2d spectrum of the real and fake quasars. (Lower panel) 1d spectrum of them. The S/N of real and fake quasars are consistent.}
    \label{fig:qso_sim}
\end{figure*}

\begin{figure*}
    \centering
    \includegraphics[width=\linewidth]{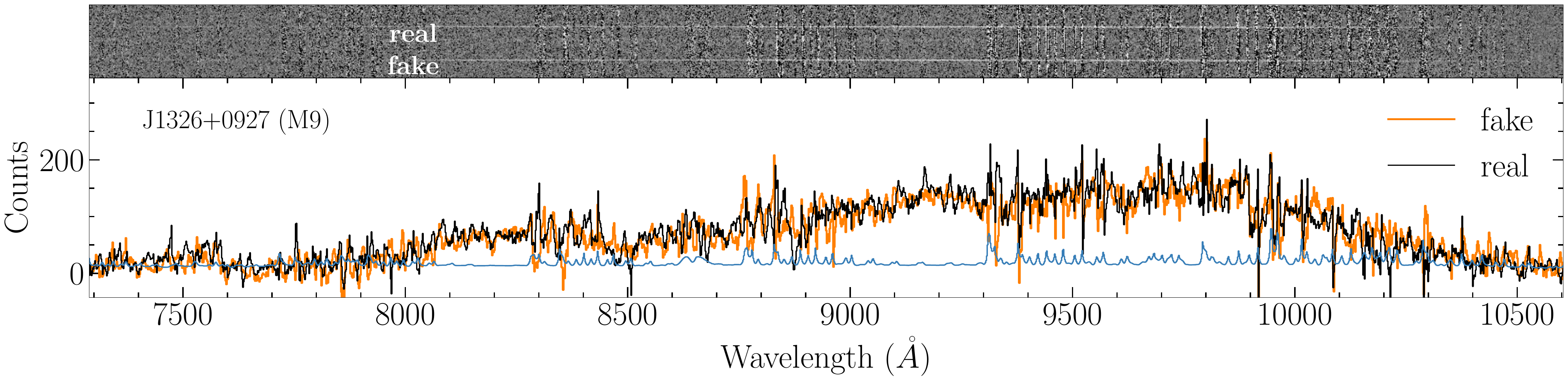}
    \caption{Simulated brown dwarf J1326+0927 in LRIS. (Upper panel) 2d spectrum of the real and fake dwarfs. (Lower panel) 1d spectrum of them. The S/N of real and fake dwarfs are consistent.}
    \label{fig:dwarf_sim}
\end{figure*}

In Figure \ref{fig:qso_sim} and Figure \ref{fig:dwarf_sim}, we show simulated observations of a quasar and a brown dwarf using LRIS. The simulated sources have the same $m_J$ and redshift (for the quasar) as the real sources in the same 2d spectra. We can see that the simulated and observed spectrum are highly consistent.

\subsection{Estimating Required Exposure Time for LRIS, MOSFIRE and NIRES}

With the simulated quasars, the next step is to define at what signal-to-noise level we can identify a quasar. We use the mean S/N of a spectrum to investigate this:

\begin{equation}
    \frac{\rm S}{\rm N}=\frac{\sum_i N_i}{\sqrt{\sum_i \sigma_i^2}}
\end{equation}

where $N_i$ represents the $N_{\rm pix}$ from the source in the $i$-th pixel, and $\sigma_i$ is the corresponding uncertainty. In practice, we use the rest-frame wavelength range $1217$-$1227\angstrom$ to calculate the mean S/N of a quasar. For the brown dwarf we use the observed-frame wavelength range $9600$-$9700\angstrom$ (for IR instruments, we use $10500$-$10600\angstrom$) to avoid the molecular absorption bands in brown dwarf spectra. 

As the minimum S/N to successfully classify a candidate, we decide to set a threshold to ${\rm S/N}=6$ based on our observing experience. With this threshold and the measured S/N of the simulated source, we use the scaling relation between exposure time and S/N to estimate the required exposure time for an unanimous classification:

\begin{equation}\label{eq:ccd}
    \frac{\rm S}{\rm N} \approx \frac{N_* t}{\sqrt{N_s t_{\rm exp}}}
\end{equation}

where $N_*$ is the count rate in photons (electrons) per second for the source of interest, $N_s$ is the pixel count rate from the sky background, and $t_{\rm exp}$ is the exposure time. We assume the noise is background dominated here, i.e. $N_s \ll N_*$.

\begin{figure*}
     \centering
     \begin{subfigure}[b]{0.32\textwidth}
         \centering
         \includegraphics[width=\textwidth]{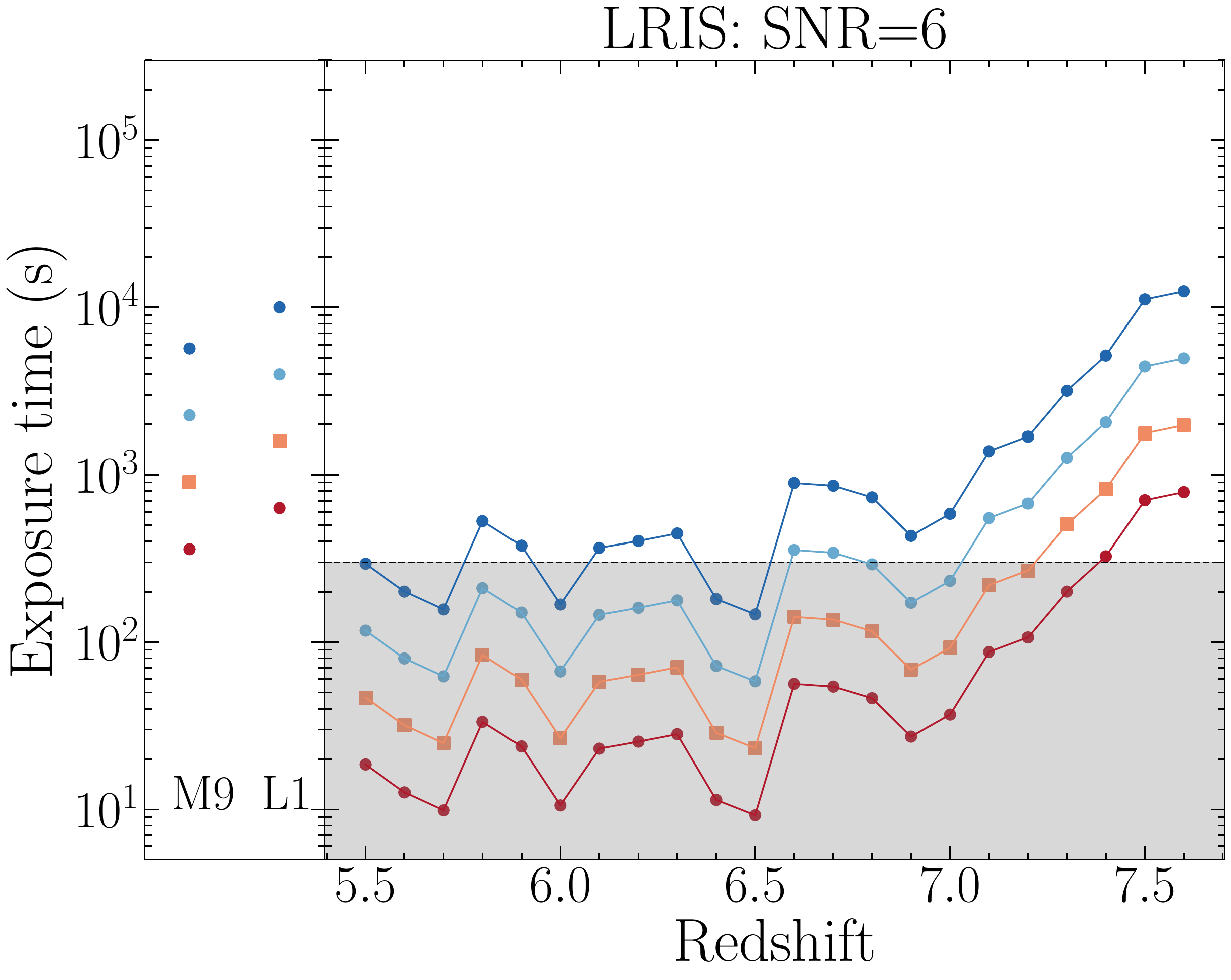}
     \end{subfigure}
     \hfill
     \begin{subfigure}[b]{0.32\textwidth}
         \centering
         \includegraphics[width=\textwidth]{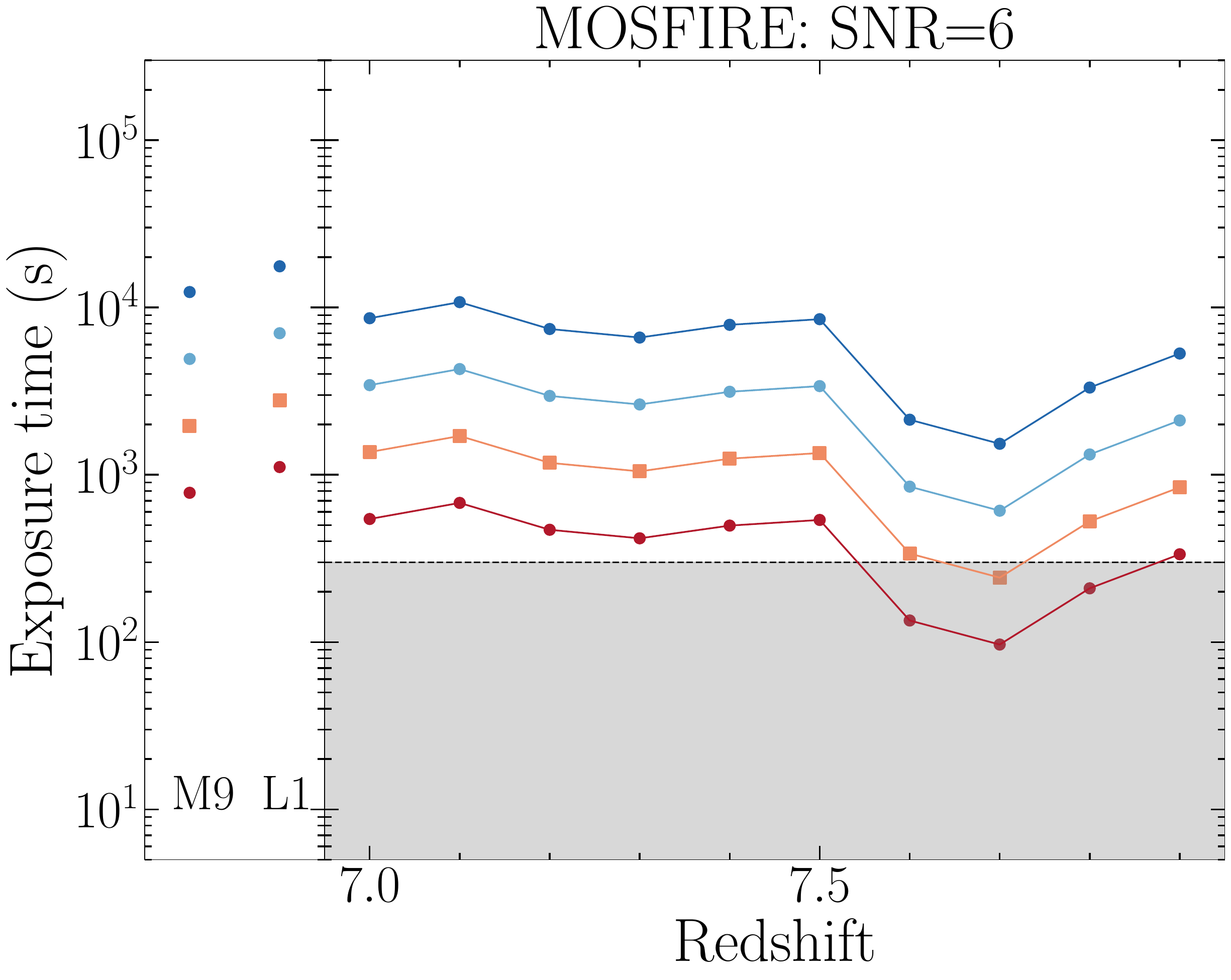}
     \end{subfigure}
     \hfill
     \begin{subfigure}[b]{0.32\textwidth}
         \centering
         \includegraphics[width=\textwidth]{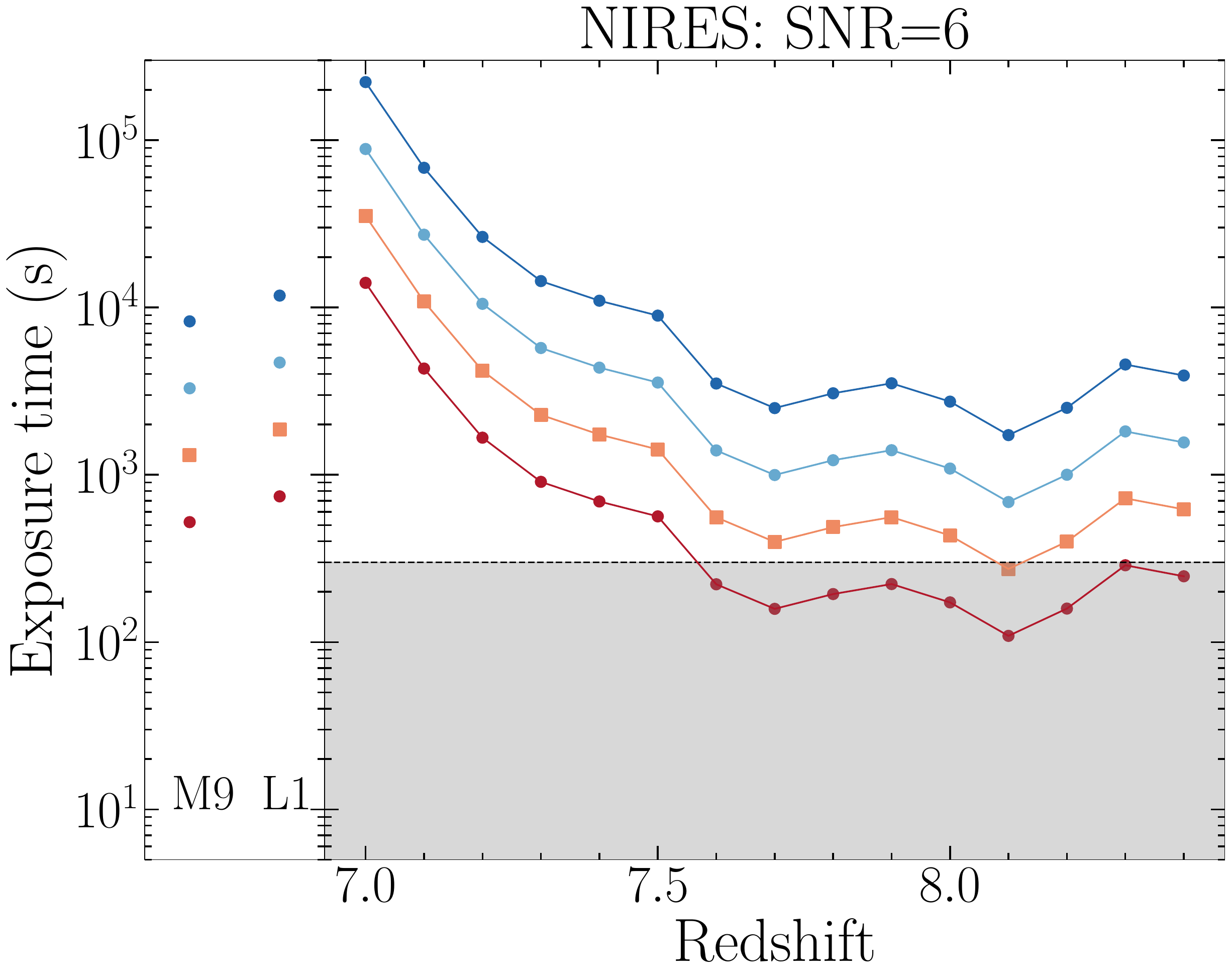}
     \end{subfigure}
     \hfill
     \begin{subfigure}[b]{0.95\textwidth}
         \centering
         \includegraphics[width=\textwidth]{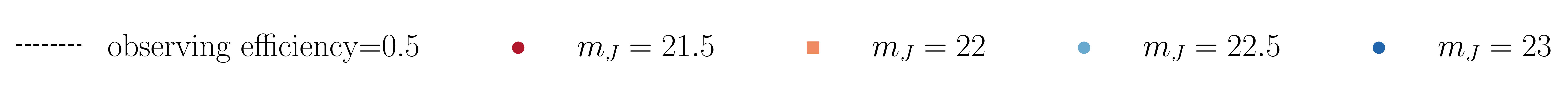}
     \end{subfigure}
        \caption{The exposure time to reach mean ${\rm S/N}=6$ for quasars with given redshift and $m_J$, as well as brown dwarfs with given spectral type and $m_J$, using (Left) LRIS, (Middle) MOSFIRE, (Right) NIRES. An observing efficiency of 0.5 indicates an exposure time of $300$s, close to the average overheads time. When the exposure time gets lower, the CCD equation will deviate from Equation \ref{eq:ccd} and at certain point the readout noise term is no longer negligible. The exposure time is therefore underestimated at the short exposure time regime.}
        \label{fig:z_exptime}
\end{figure*}

Now that we can simulate a quasar with arbitrary redshift $z$ and apparent magnitude $m_J$, and have defined that we can identify it with ${\rm S/N}\geq 6$, we use Equation \ref{eq:ccd} to estimate the required exposure time to identify this quasar with LRIS, MORSFIRE or NIRES. We show the results in Figure \ref{fig:z_exptime}. As a comparison, we also show the required exposure times for brown dwarfs with different $m_J$ in the same figure. The average overheads time is $\sim 300$s, thus an observing efficiency of 0.5 in the plot indicates exposing for $300$s as well. The shape of the $z$-$t_{\rm exp}$ curves for quasars is generally determined by the throughput of the spectrometer and the telluric absorption bands. 

\begin{figure}
    \centering
    \includegraphics[width=\columnwidth]{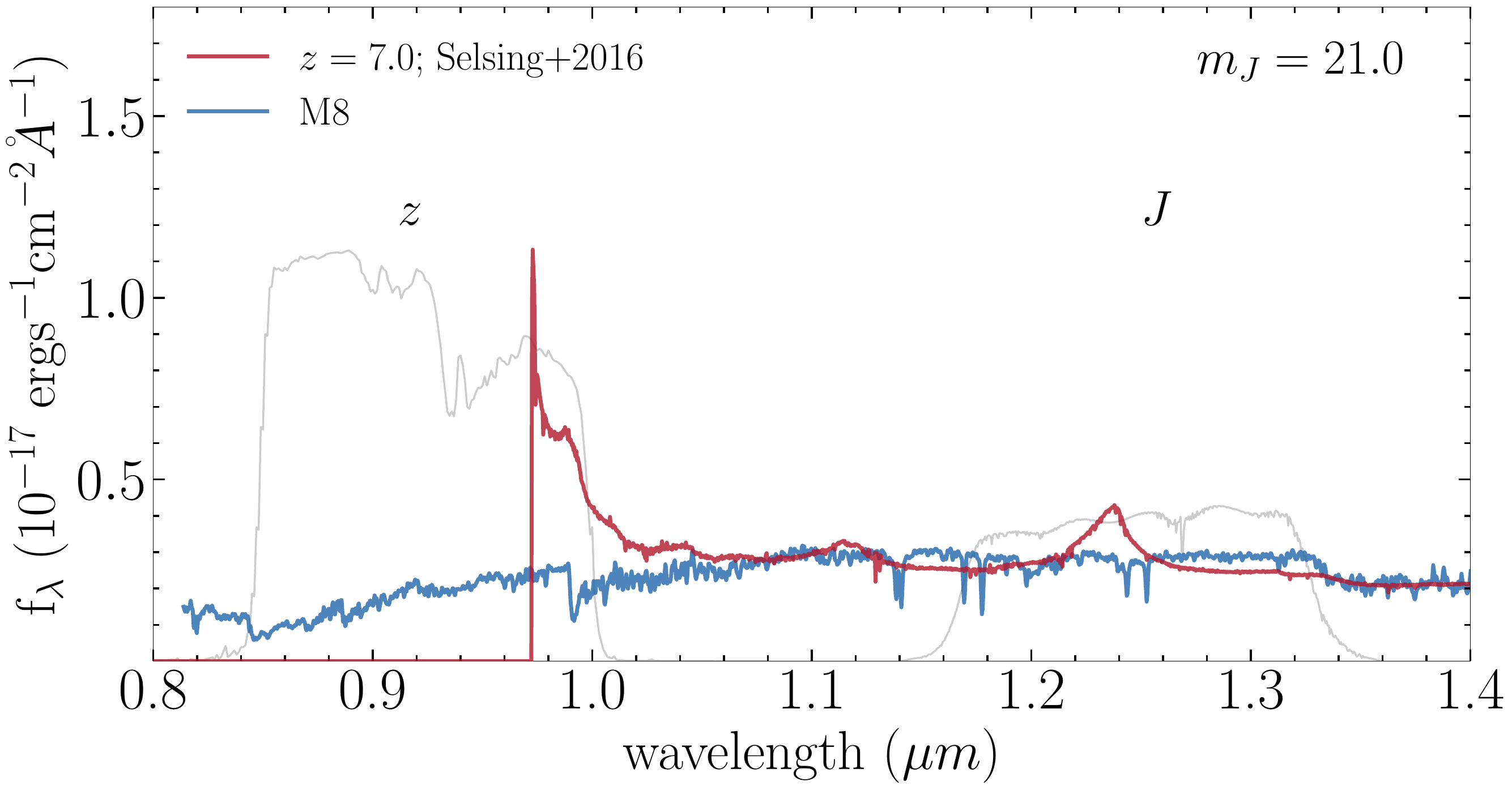}
    \caption{The spectrum of quasar and M8 dwarf with $m_{\rm J}=21.0$ and a similar $z$-band magnitudes. We can see that the quasar will look much brighter than the brown dwarf in LRIS working wavelength, explaining the relative short exposure times to identify a quasar.}
    \label{fig:qso_star}
\end{figure}

For bright quasars, the exposure time can be surprisingly low, while stars with similar apparent magnitude require much longer exposure time. This is consistent with the fact that the observed quasars are generally much brighter than the contaminants. In Figure \ref{fig:qso_star} we show a quasar spectrum and a brown dwarf spectrum with the same $m_J$. They are similarly bright in the detection band ($J$-band) and drop-out band ($z$-band), but the quasar fluxes are concentrated in a much smaller wavelength range, and therefore produce a higher signal-to-noise detection in the LRIS spectrum.

With Figure \ref{fig:z_exptime}, we also confirm that we can identify quasars up to $z\sim 7.6$ with LRIS. We can identify a $z\sim 7.5$, $m_{\rm J}=21.5$ quasar with a $\sim 15$ min exposure. It is also possible to identify faint quasars that are common in \textit{Euclid} with LRIS. We can confirm a $z\sim 7.0$, $m_{\rm J}=23.0$ quasar around $15$ min, a $z\sim 7.5$, $m_{\rm J}=23.0$ quasar within 3 hours.

\begin{figure}
    \centering
    \includegraphics[width=0.96\columnwidth]{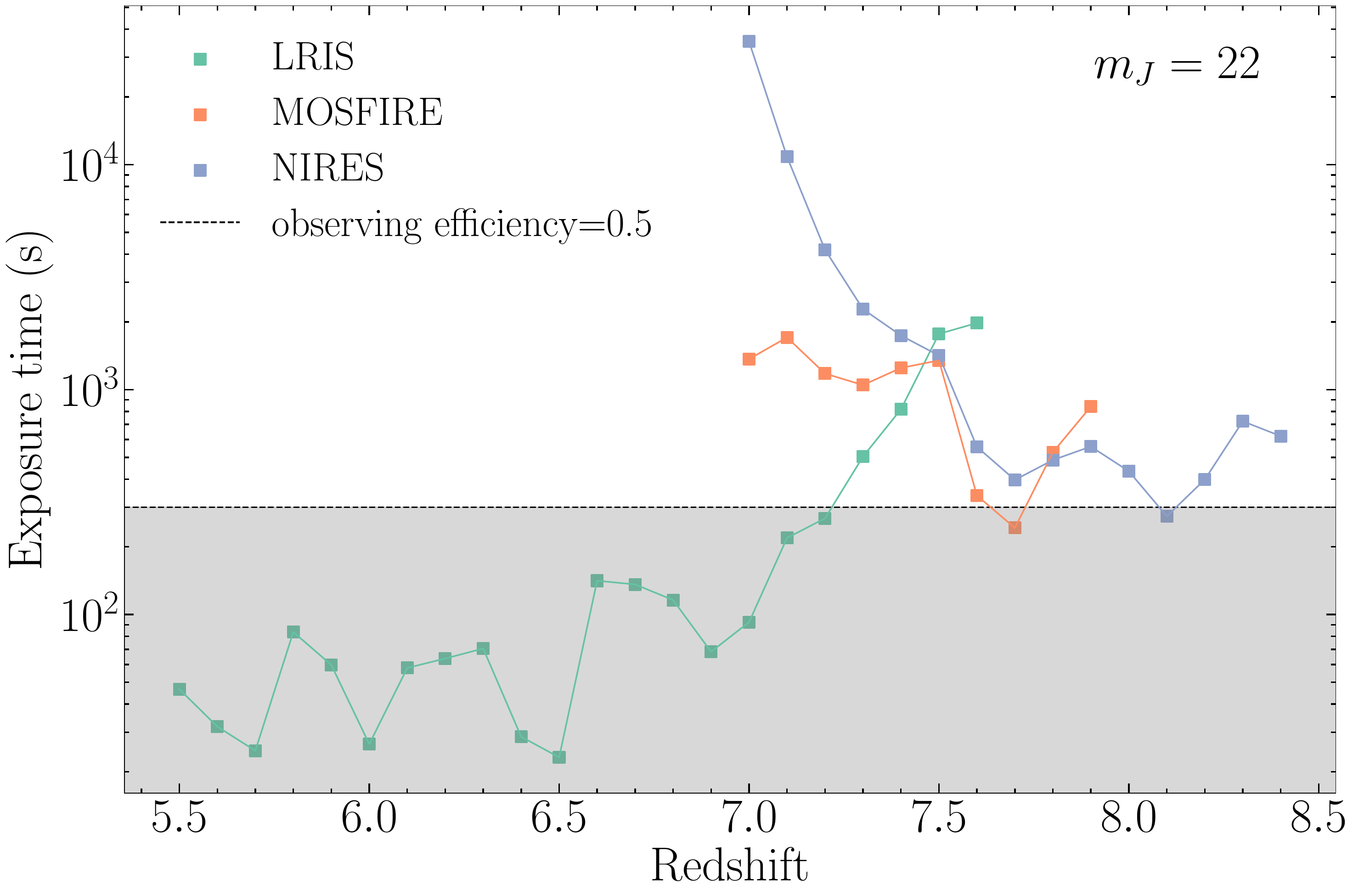}
    \caption{The exposure time to reach mean ${\rm S/N}=6$ for quasars with $m_J=22$ and given redshift, observed with three different instruments.}
    \label{fig:z_exptime_all}
\end{figure}

In Figure \ref{fig:z_exptime_all}, we combine the results of all three instruments. From these figures we can have a clearer idea of the optimal observing strategy: for $z<7.5$ quasar candidates, LRIS is the most suitable instrument; for candidates with higher estimated redshifts, IR instruments like MOSFIRE or NIRES are required. Alternatively, we can use LRIS for all the candidates first, and use the other two instruments to observe the candidates with no LRIS fluxes.

The exposure time calculators provided by Keck\footnote{The calculator for LRIS is available at: \href{http://etc.ucolick.org/web\_s2n/lris}{http://etc.ucolick.org/web\_s2n/lris}, for MOSFIRE at: \href{https://www2.keck.hawaii.edu/inst/mosfire/etc.html}{https://www2.keck.hawaii.edu/inst/mosfire/etc.html}.} are not well-tested for the purpose of quasar confirmation. Our simulation with noise maps from real observations should be more robust in practice. Furthermore, the visualization of the 2d spectrum is highly informative to support classifications for future quasar search campaigns. We also note that this simulation tool which is based on the \texttt{PypeIt} routine can be useful to simulate other kinds of sources.

\section{Summary}

In this paper, we present the spectroscopic observations of 174 $z\gtrsim 5$ quasar candidates from various quasar searches. We reduce the spectra taken with Keck/LRIS, Keck/MOSFIRE, and Keck/NIRES using the \texttt{PypeIt} package.
Based on the resulting 2d and 1d spectra, we classify sources into quasars and brown dwarfs where possible. The main results based on these data are summarised below:

\begin{enumerate}
\item We identify seven $z\sim 6$ quasars (QSO), three of which were selected with \texttt{XDHZQSO} \citep{Nanni2022} using VIKING/UKIDSS and are newly reported in this paper.
We also classify 51 candidates as dwarf stars with spectral types M, L, and T (STAR). 
The spectra of 74 candidates do not resemble quasars, but cannot be further classified (UNQ) due to the low signal-to-noise ratio of the spectra.
Two candidates have low signal-to-noise ratio spectra and/or too short wavelength coverage. As they could possibly be high-$z$ quasars we label them inconclusive (INCONCLUSIVE).
For 42 sources we could not fully reduce and/or extract the spectroscopy. These candidates have no classification and are labeled FAIL.
\item Based on the classifications, we show the distribution of all the sources in the color space and relative flux space with photometry measurements from DELS ($z$), VIKING/UKIDSS ($YJHK_s$/$YJHK$), and un\textit{WISE} ($W1W2$) (Figure \ref{fig:flux_ratio_corner}). 
The brown dwarfs in our samples inhabit the high-$z$ quasar locus, and deviate from the general brown dwarf distribution implied by an empirical brown dwarf model, which is adopted in various candidate selection methods. 
We estimate the probability that the confirmed brown dwarfs are consistent with the brown dwarf model (Section \ref{sec:result_relflux}). 
Our order of magnitude estimate suggests that the model has difficulty to explain the observed brown dwarf distribution driven mainly by discrepancies in the $K$-band and $W1$-band photometry. 
The non-Gaussian noise of the photometric observation or the variability of the brown dwarfs could explain these discrepancies. 
To mitigate the non-Gaussian effects on the photometric noise, we propose careful visual inspection of the photometric images or follow-up photometric observations. 
To capture the effects of variable sources, we suggest to adopt a single, empirical contaminant model, or to refine the brown dwarf models to include these observed outliers.
\item We simulate the spectroscopic observations of high-$z$ quasars. Based on the synthetic spectroscopy, we estimate the required exposure time to identify a quasar with a given redshift and $J$-band magnitude using Keck/LRIS, Keck/MOSFIRE and Keck/NIRES. We prove the feasibility of using LRIS to identify $z<7.6$ quasars and quasars as faint as $m_J\sim 23$. We also find that the required exposure time to identify a quasar with LRIS is generally smaller than the time to identify a brown dwarf with the same apparent magnitude, which is due to the difference between their spectral shapes. 
We compare the required exposure time to reach a signal-to-noise ratio of 6 for a $m_{J}=22$ quasars at different redshifts using LRIS, MOSFIRE, and NIRES (Figure \ref{fig:z_exptime_all}). We conclude that the optimal strategy is to use LRIS to identify $z<7.5$ quasar candidates, and to utilize the other two IR instruments for candidates with higher estimated redshifts. 
\end{enumerate}


\section*{Acknowledgements}

We thank Suk Sien Tie, Silvia Onorato, and Ben Wang who supported this project during observing runs. We also thank the ENIGMA group members at Leiden University and UCSB for providing valuable comments on the manuscript. 

JTS, RN, and JFH acknowledge funding from the European Research Council (ERC) Advanced Grant program under the European Union’s Horizon 2020 research and innovation programme (Grant agreement No. 885301). 

The data presented herein were obtained at the W. M. Keck Observatory, which is operated as a scientific partnership among the California Institute of Technology, the University of California and the National Aeronautics and Space Administration. The Observatory was made possible by the generous financial support of the W. M. Keck Foundation. This research has also made use of the Keck Observatory Archive (KOA), which is operated by the W. M. Keck Observatory and the NASA Exoplanet Science Institute (NExScI), under contract with the National Aeronautics and Space Administration.

\textit{Software:} \texttt{Astropy}\citep[][]{astropy}, \texttt{PypeIt} \citep[][]{pypeit:joss_pub}, \texttt{simqso} \citep[][]{McGreer2013}, \texttt{scikit-learn} \citep[][]{scikit-learn}, \texttt{NumPy} \citep[][]{numpy}, \texttt{Matplotlib} \citep[][]{matplotlib}.


\section*{Data Availability}

The HzQCA catalog is published in the electronic version of the journal. The scripts to reduce all the spectroscopic data and reproduce all the results in this paper are stored in a Github repository\footnote{\href{https://github.com/enigma-igm/highz\_qso\_arxiv}{https://github.com/enigma-igm/highz\_qso\_arxiv}}. All the data are available from the corresponding author upon reasonable request. This work uses publicly available data from the UKIDSS DR11, VIKING DR4, the Pan-STARRS, and the DELS surveys.



\bibliographystyle{mnras}
\bibliography{ref} 




\appendix

\section{Reduced Spectra}\label{sec:spectrum}


\begin{figure*}
    \centering
    \includegraphics[width=\linewidth]{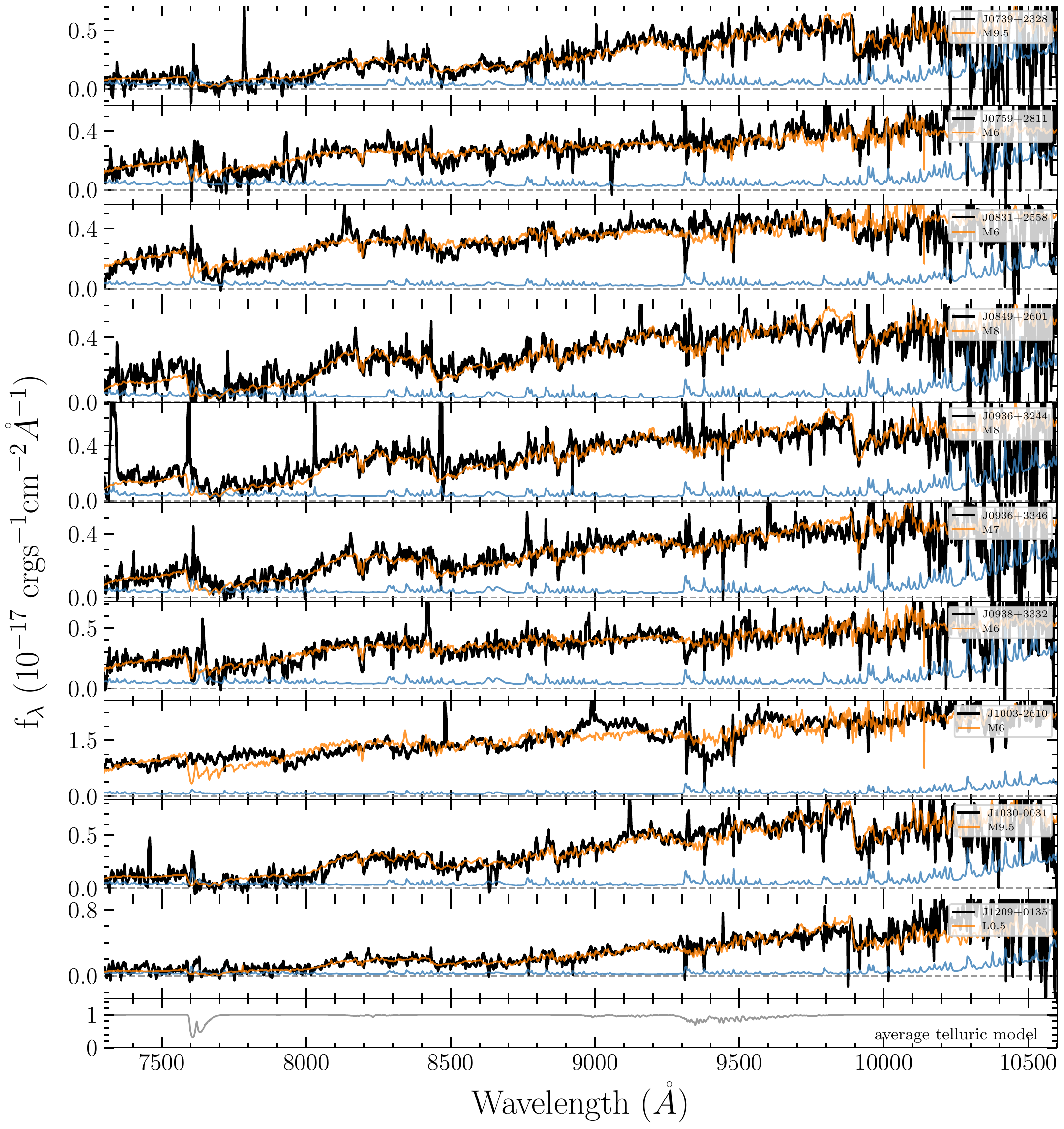}
    \caption{Final 1d spectra of STAR targets in LRIS runs. The 1d spectra here are fluxed and telluric corrected, and the orange curves are the brown dwarf spectra with the lowest reduced $\chi^2$ values. The blue curves are the noise vectors. The grey curve in the bottom panel is the average telluric model. The spectra and noise vectors are smoothed using the inverse variance with a smoothing window of 5.}
    \label{fig:lris_star}
\end{figure*}

\begin{figure*}
    \centering
    \includegraphics[width=\linewidth]{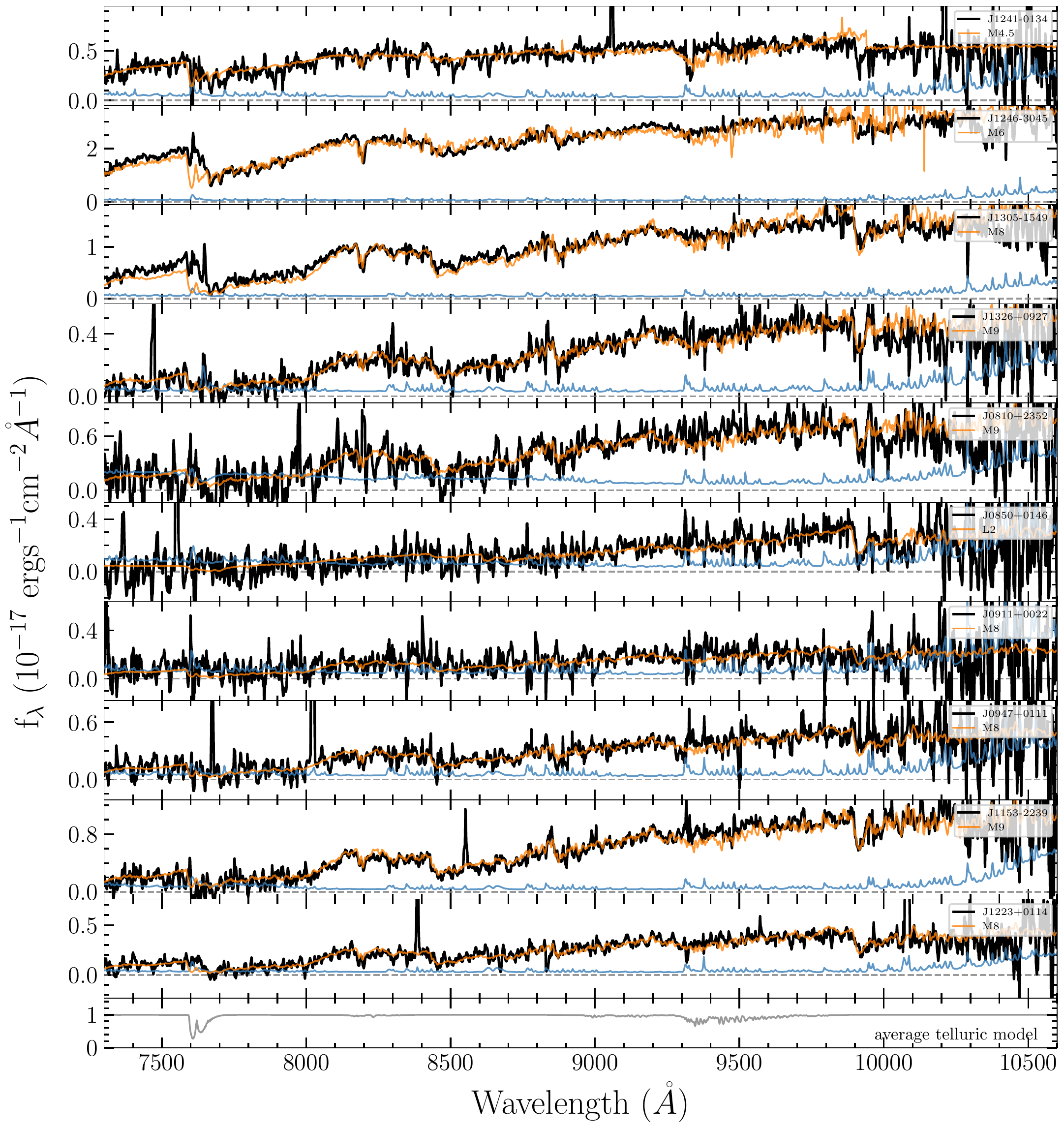}
    \contcaption{Final 1d spectra of STAR targets in LRIS runs. The 1d spectra here are fluxed and telluric corrected, and the orange curves are the brown dwarf spectra with the lowest reduced $\chi^2$ values. The blue curves are the noise vectors. The grey curve in the bottom panel is the average telluric model. The spectra and noise vectors are smoothed using the inverse variance with a smoothing window of 5.}
\end{figure*}

\begin{figure*}
    \centering
    \includegraphics[width=\linewidth]{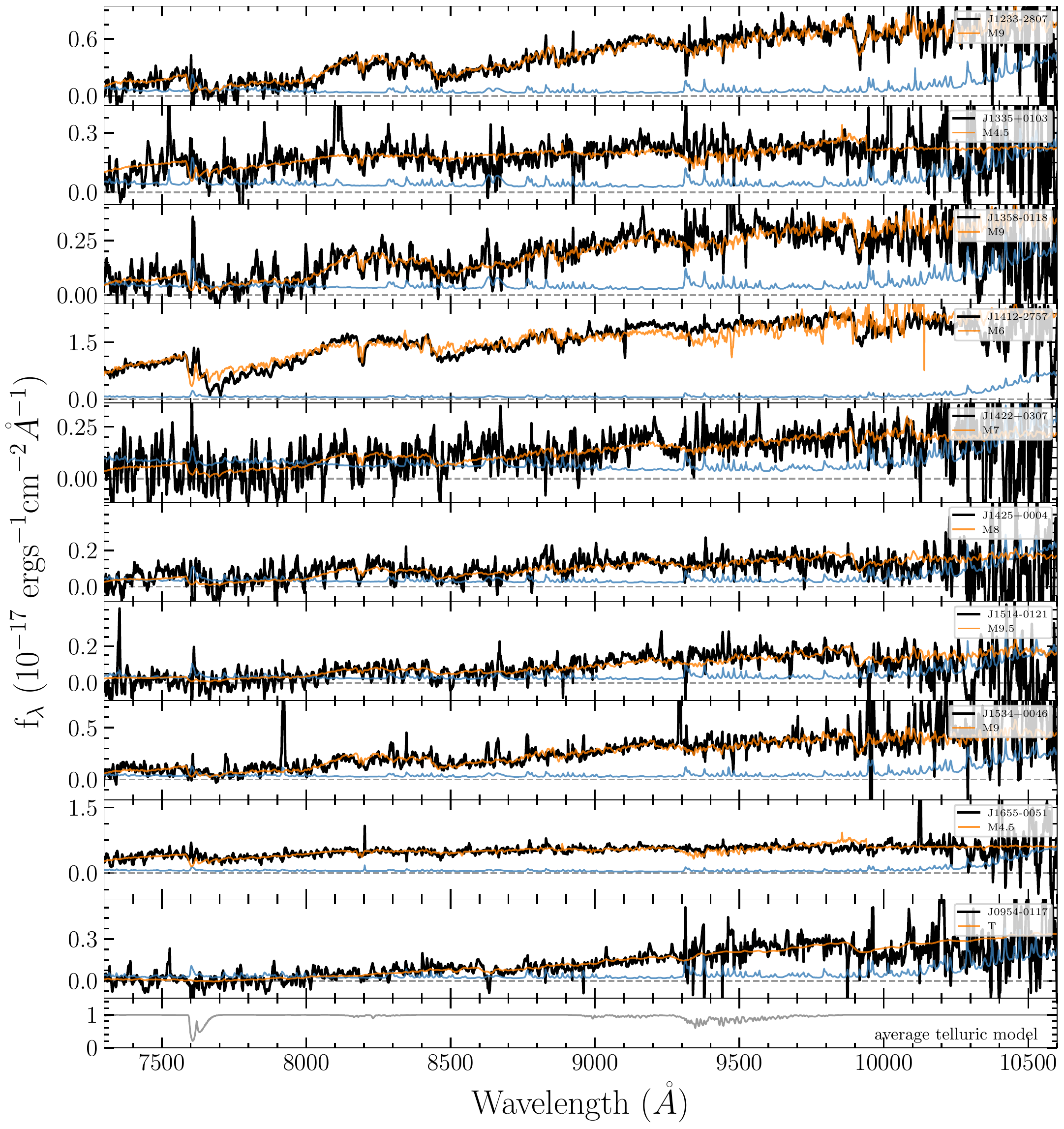}
    \contcaption{Final 1d spectra of STAR targets in LRIS runs. The 1d spectra here are fluxed and telluric corrected, and the orange curves are the brown dwarf spectra with the lowest reduced $\chi^2$ values. The blue curves are the noise vectors. The grey curve in the bottom panel is the average telluric model. The spectra and noise vectors are smoothed using the inverse variance with a smoothing window of 5.}
\end{figure*}

\begin{figure*}
    \centering
    \includegraphics[width=\linewidth]{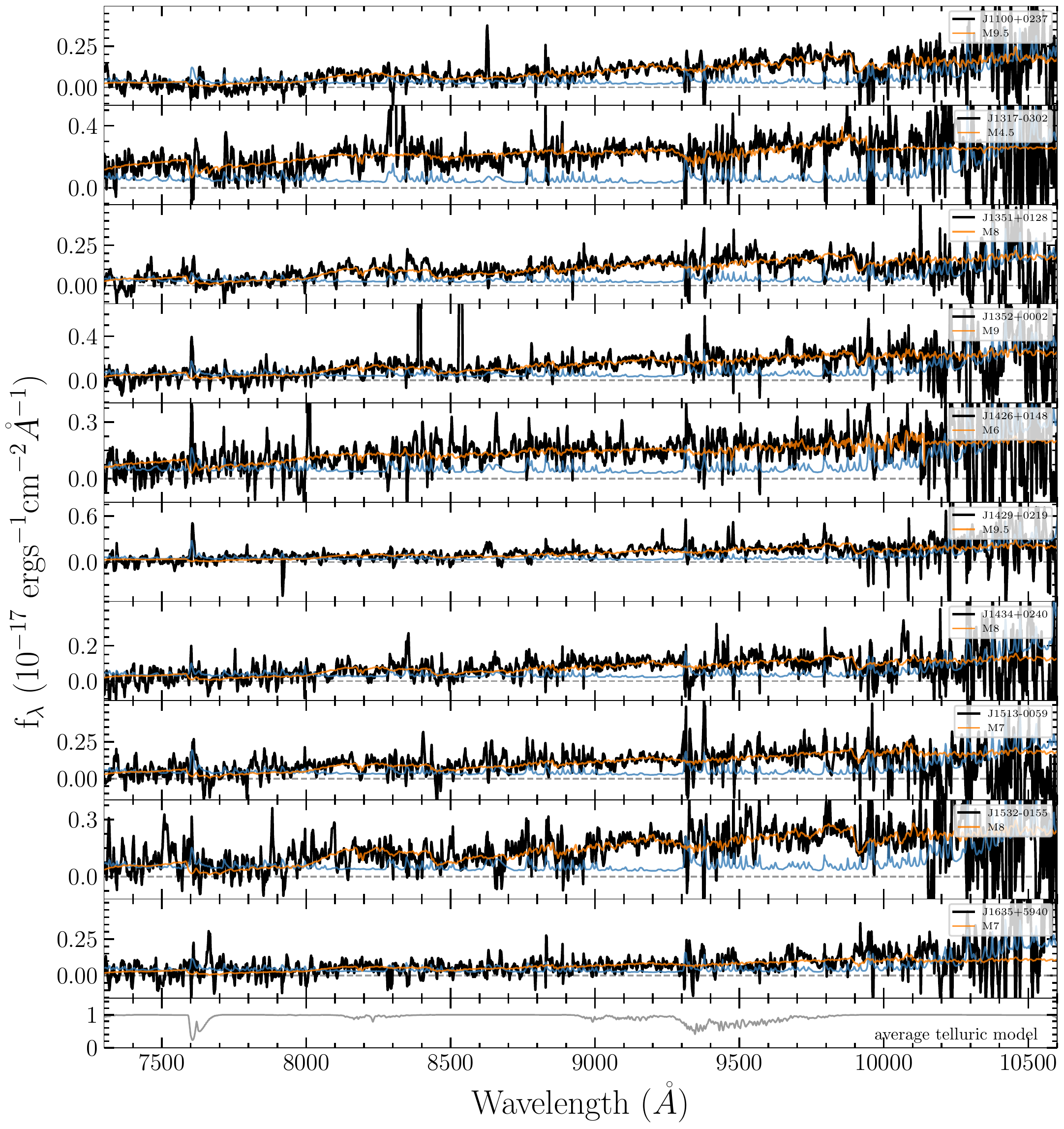}
    \contcaption{Final 1d spectra of STAR targets in LRIS runs. The 1d spectra here are fluxed and telluric corrected, and the orange curves are the brown dwarf spectra with the lowest reduced $\chi^2$ values. The blue curves are the noise vectors. The grey curve in the bottom panel is the average telluric model. The spectra and noise vectors are smoothed using the inverse variance with a smoothing window of 5.}
\end{figure*}

\begin{figure*}
    \centering
    \includegraphics[width=\linewidth]{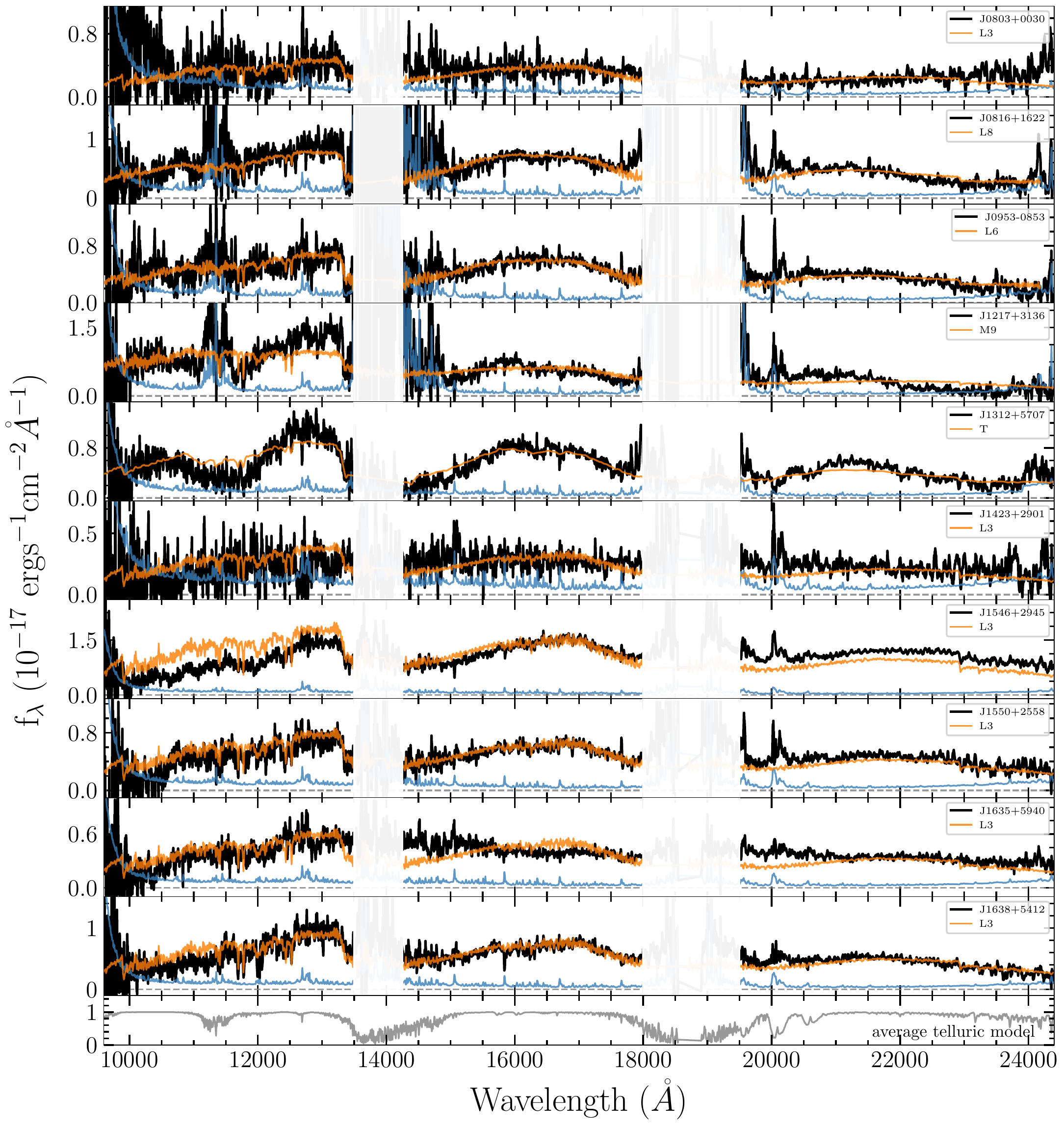}
    \caption{Final 1d spectra of STAR targets in NIRES runs. The 1d spectra here are fluxed and telluric corrected, and the orange curves are the brown dwarf spectra with the lowest reduced $\chi^2$ values. The blue curves are the noise vectors. The grey curve in the bottom panel is the average telluric model. The spectra and noise vectors are smoothed using the inverse variance with a smoothing window of 5.}
    \label{fig:nires_star}
\end{figure*}

\begin{figure*}
    \centering
    \includegraphics[width=\linewidth]{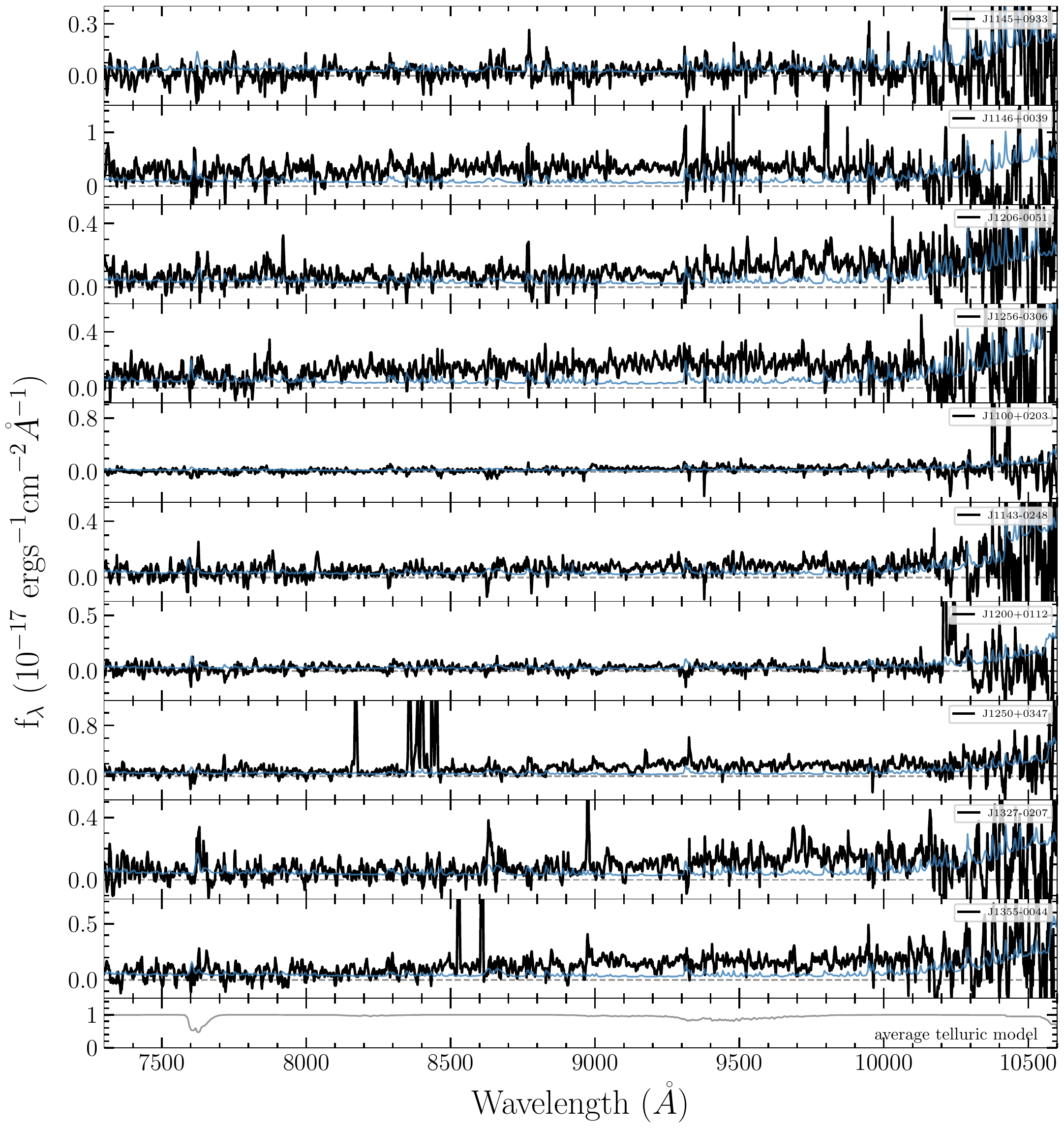}
    \caption{Final 1d spectra of UNQ targets in LRIS runs. The 1d spectra here are fluxed and telluric corrected. The blue curves are the noise vectors. The grey curve in the bottom panel is the average telluric model. The spectra and noise vectors are smoothed using the inverse variance with a smoothing window of 5.}
    \label{fig:lris_uNQ}
\end{figure*}

\begin{figure*}
    \centering
    \includegraphics[width=\linewidth]{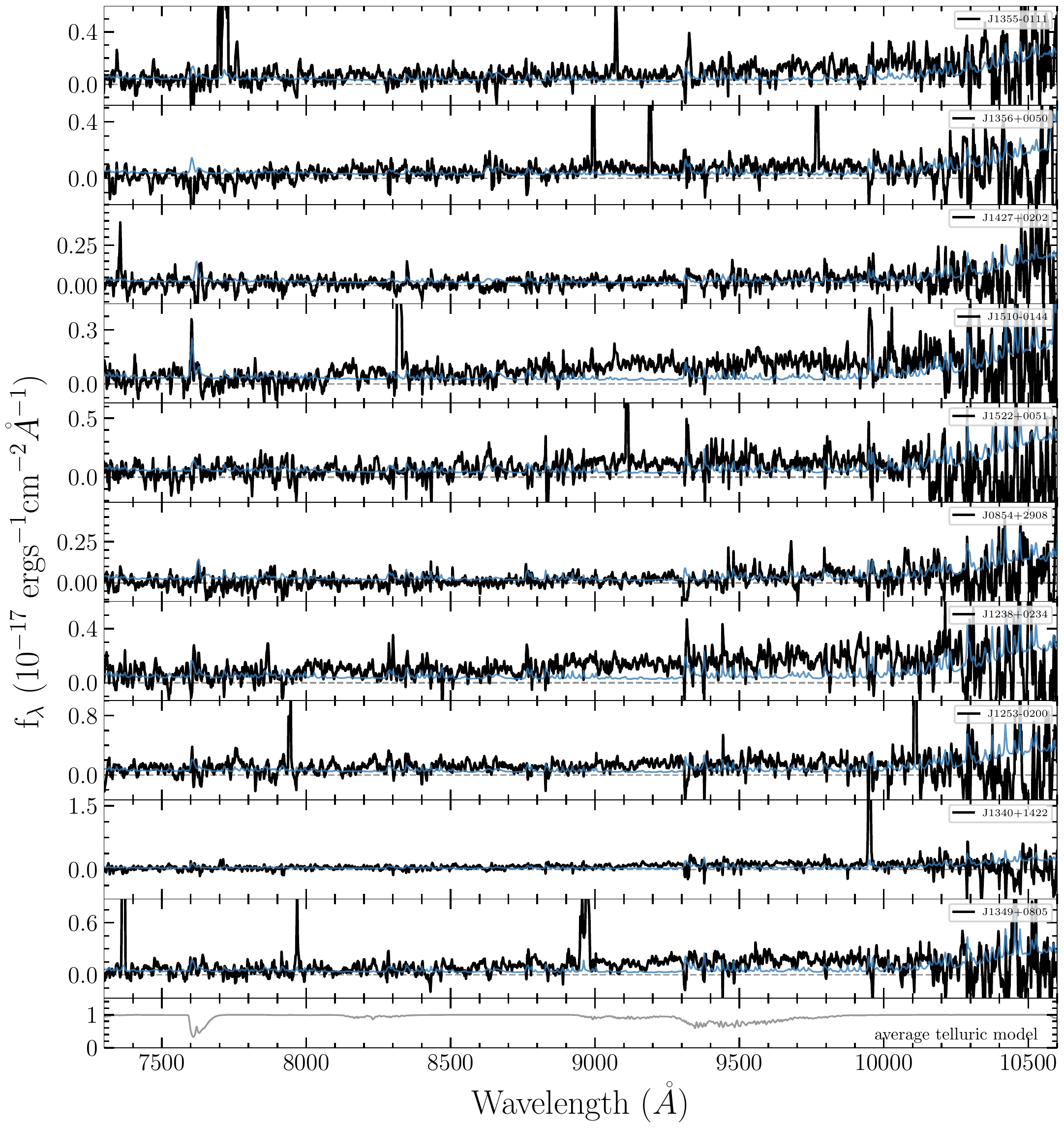}
    \contcaption{Final 1d spectra of UNQ targets in LRIS runs. The 1d spectra here are fluxed and telluric corrected. The blue curves are the noise vectors. The grey curve in the bottom panel is the average telluric model. The spectra and noise vectors are smoothed using the inverse variance with a smoothing window of 5.}
\end{figure*}

\begin{figure*}
    \centering
    \includegraphics[width=\linewidth]{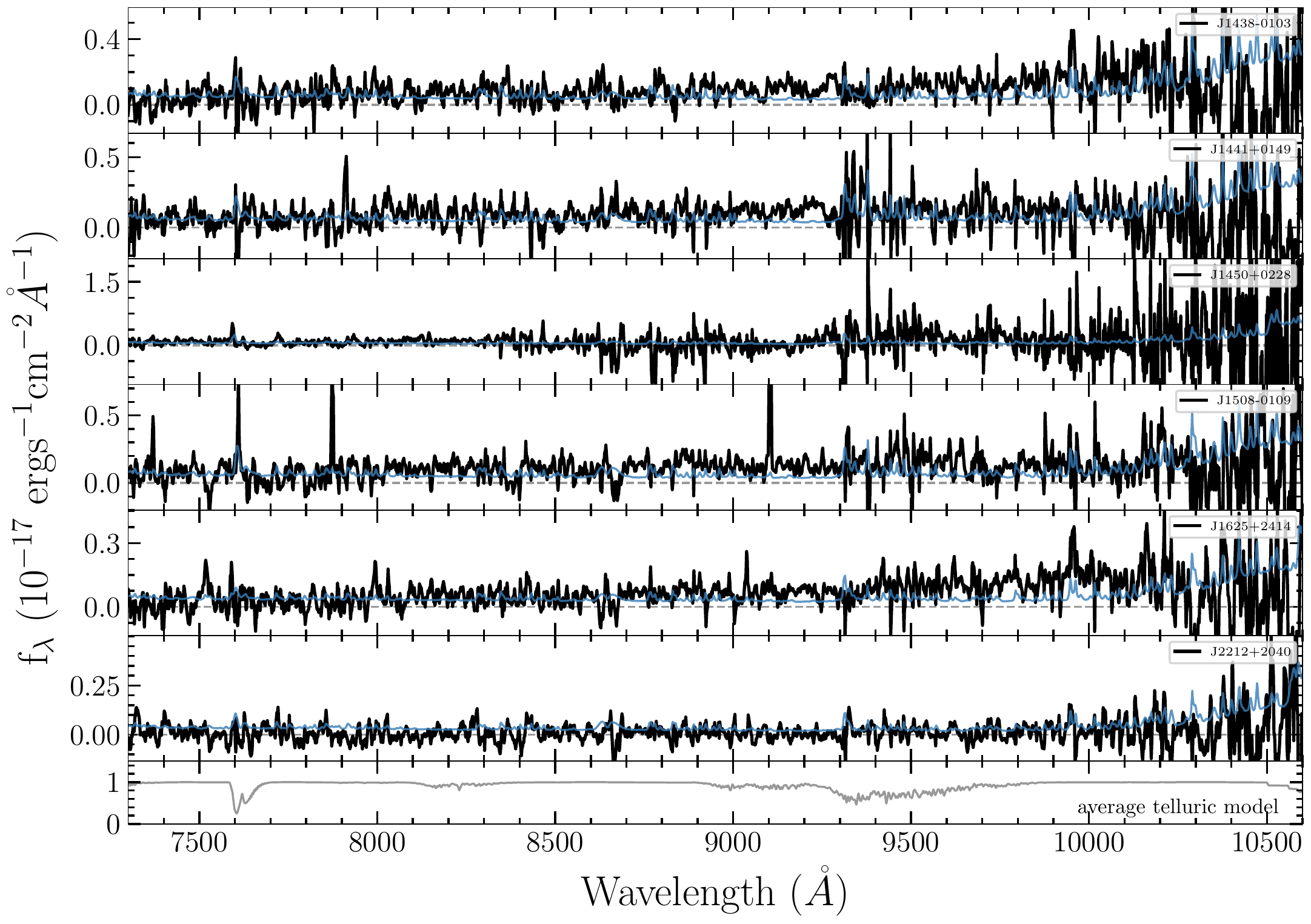}
    \contcaption{Final 1d spectra of UNQ targets in LRIS runs. The 1d spectra here are fluxed and telluric corrected. The blue curves are the noise vectors. The grey curve in the bottom panel is the average telluric model. The spectra and noise vectors are smoothed using the inverse variance with a smoothing window of 5.}
\end{figure*}

\begin{figure*}
    \centering
    \includegraphics[width=\linewidth]{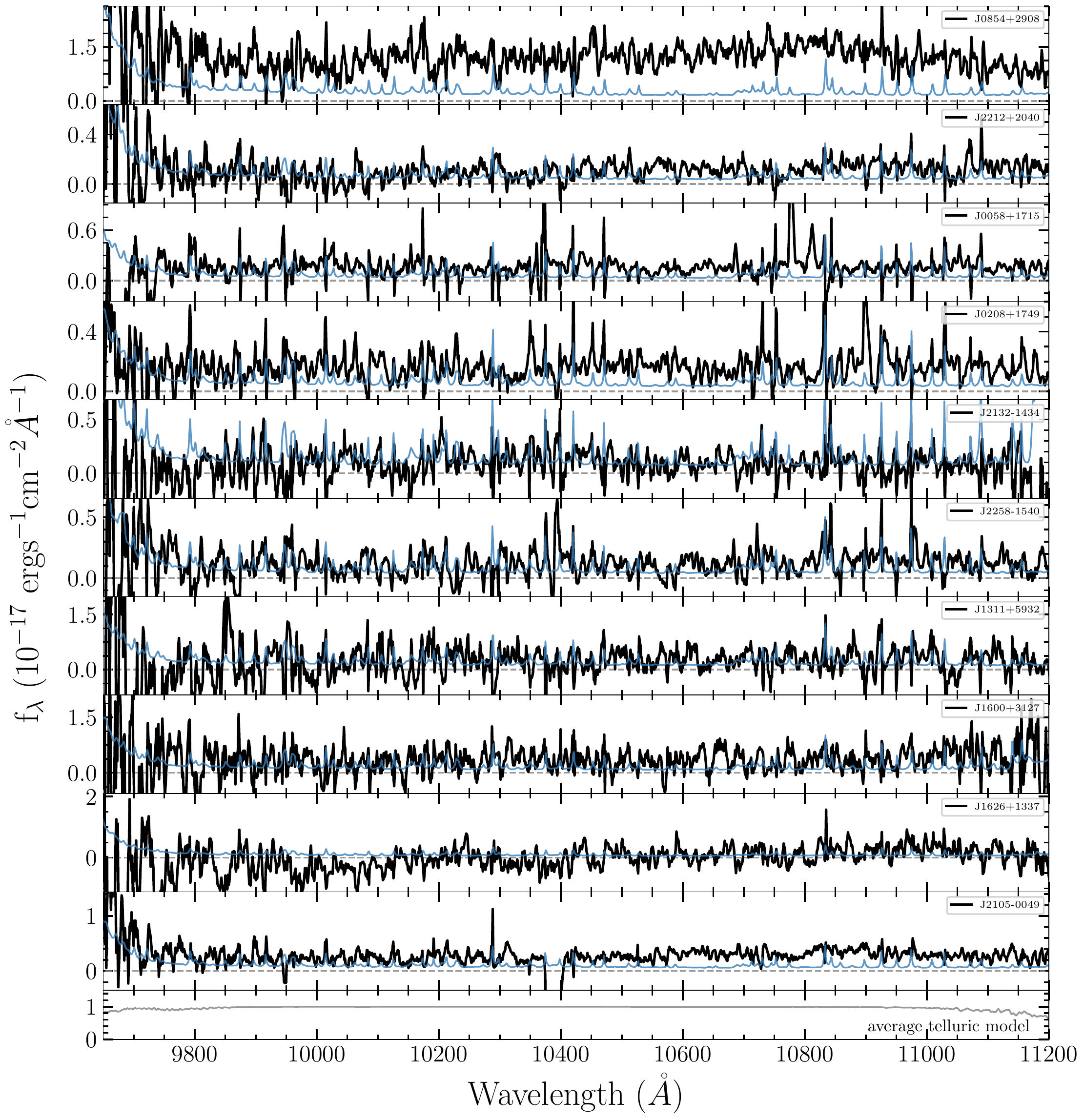}
    \caption{Final 1d spectra of UNQ targets in MOSFIRE runs. The 1d spectra here are fluxed and telluric corrected. The blue curves are the noise vectors. The grey curve in the bottom panel is the average telluric model. The spectra and noise vectors are smoothed using the inverse variance with a smoothing window of 5.}
    \label{fig:mosfire_uNQ}
\end{figure*}

\begin{figure*}
    \centering
    \includegraphics[width=\linewidth]{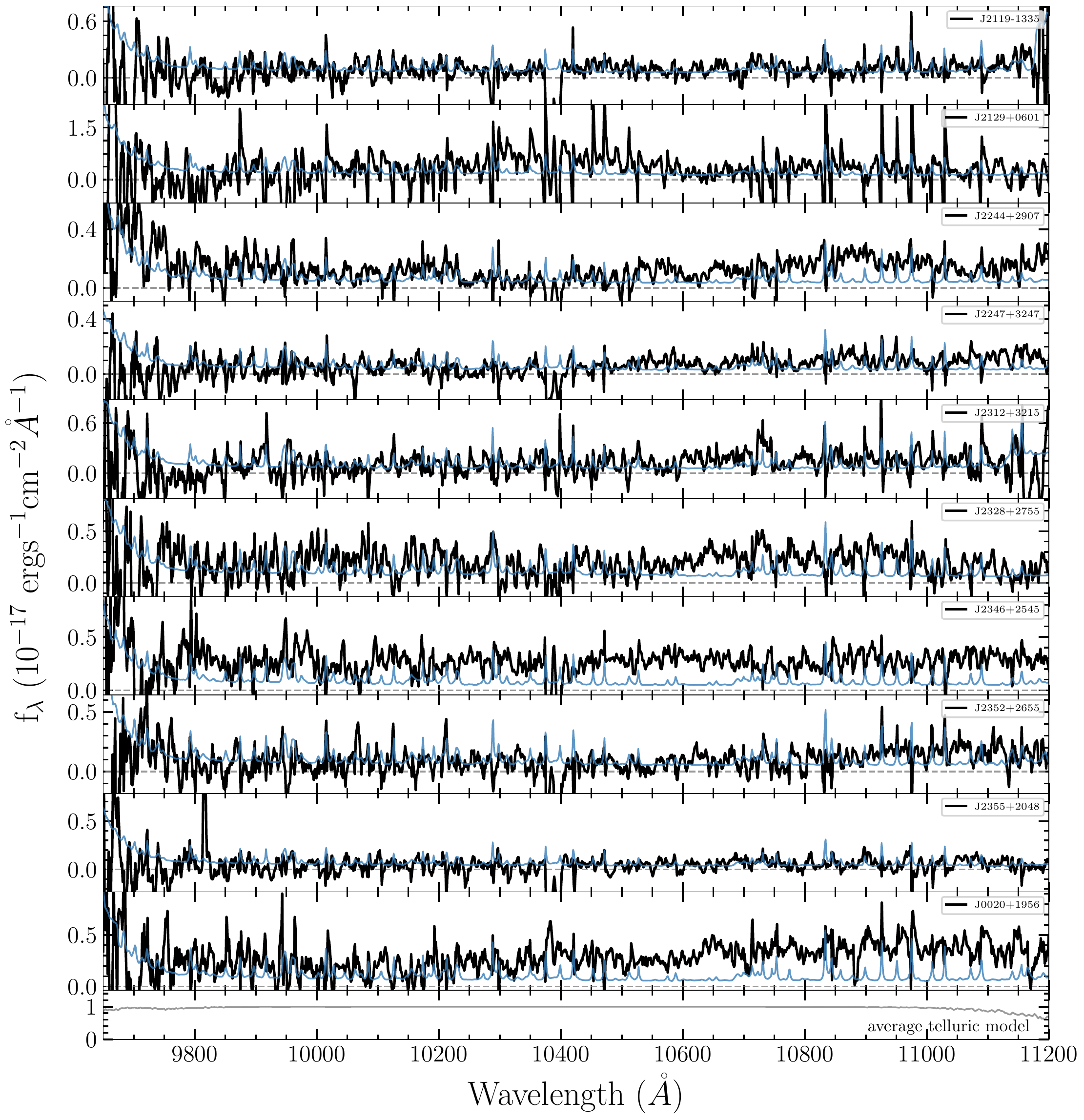}
    \contcaption{Final 1d spectra of UNQ targets in MOSFIRE runs. The 1d spectra here are fluxed and telluric corrected. The blue curves are the noise vectors. The grey curve in the bottom panel is the average telluric model. The spectra and noise vectors are smoothed using the inverse variance with a smoothing window of 5.}
\end{figure*}

\begin{figure*}
    \centering
    \includegraphics[width=\linewidth]{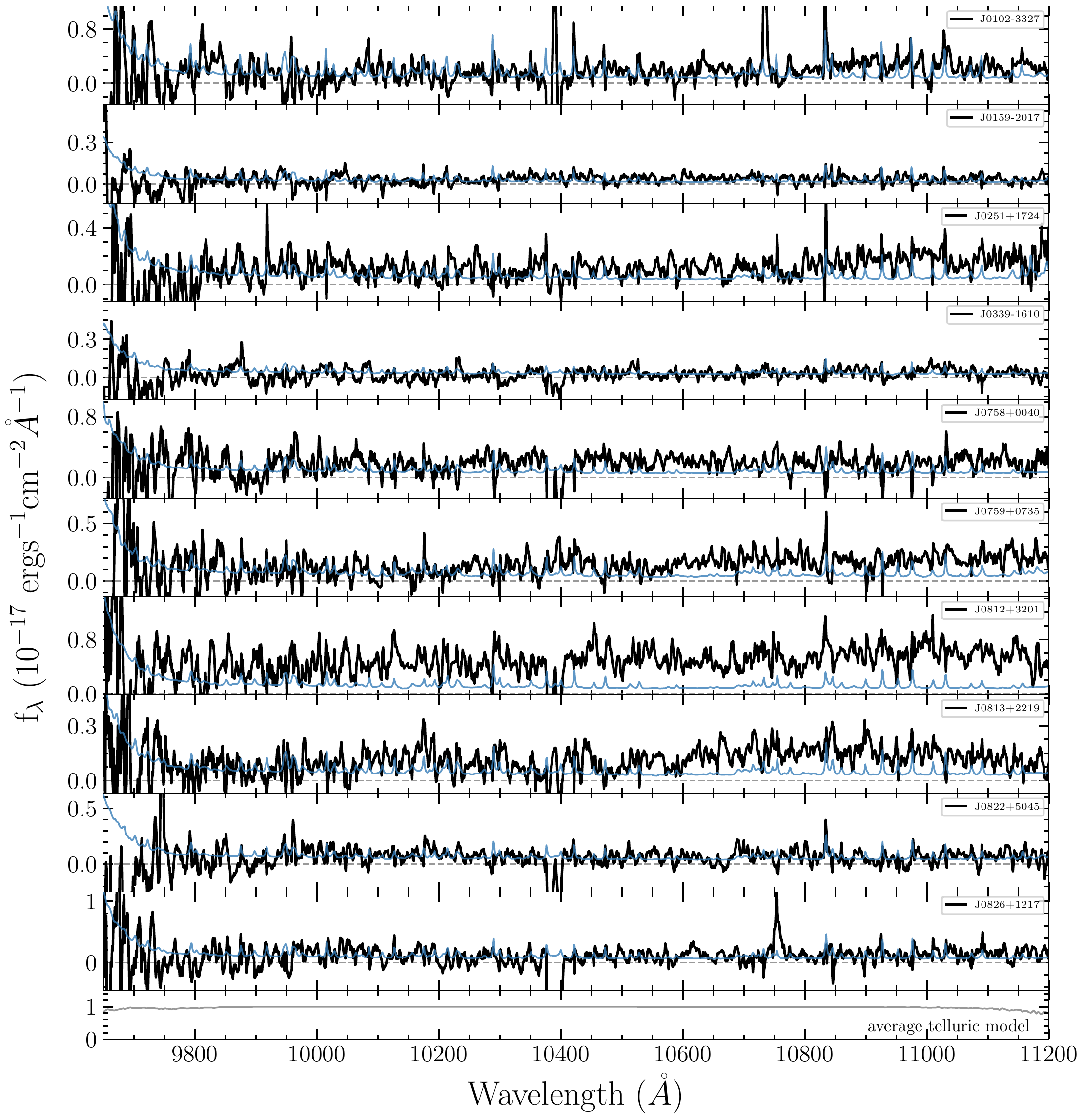}
    \contcaption{Final 1d spectra of UNQ targets in MOSFIRE runs. The 1d spectra here are fluxed and telluric corrected. The blue curves are the noise vectors. The grey curve in the bottom panel is the average telluric model. The spectra and noise vectors are smoothed using the inverse variance with a smoothing window of 5.}
\end{figure*}

\begin{figure*}
    \centering
    \includegraphics[width=\linewidth]{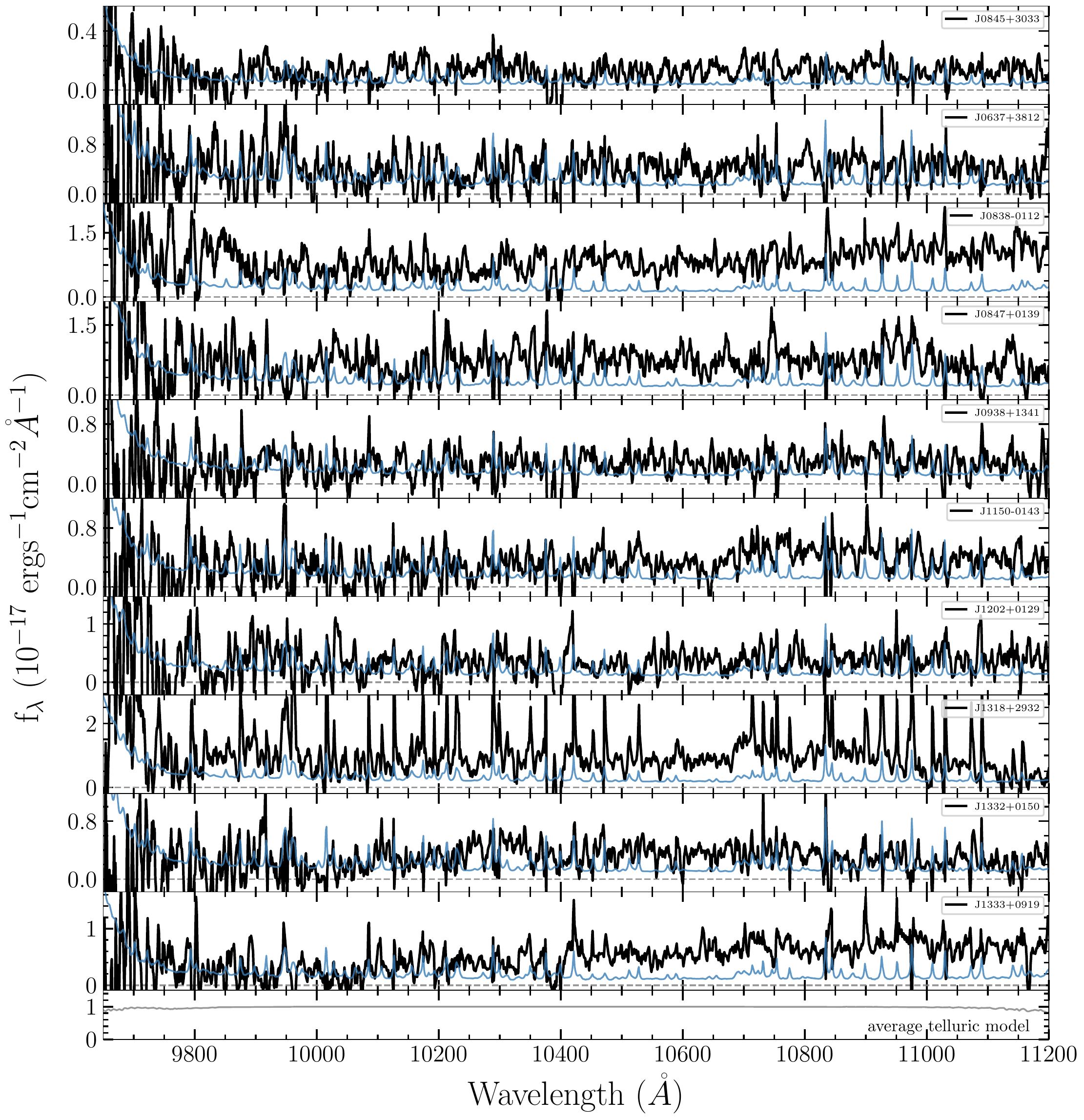}
    \contcaption{Final 1d spectra of UNQ targets in MOSFIRE runs. The 1d spectra here are fluxed and telluric corrected. The blue curves are the noise vectors. The grey curve in the bottom panel is the average telluric model. The spectra and noise vectors are smoothed using the inverse variance with a smoothing window of 5.}
\end{figure*}

\begin{figure*}
    \centering
    \includegraphics[width=\linewidth]{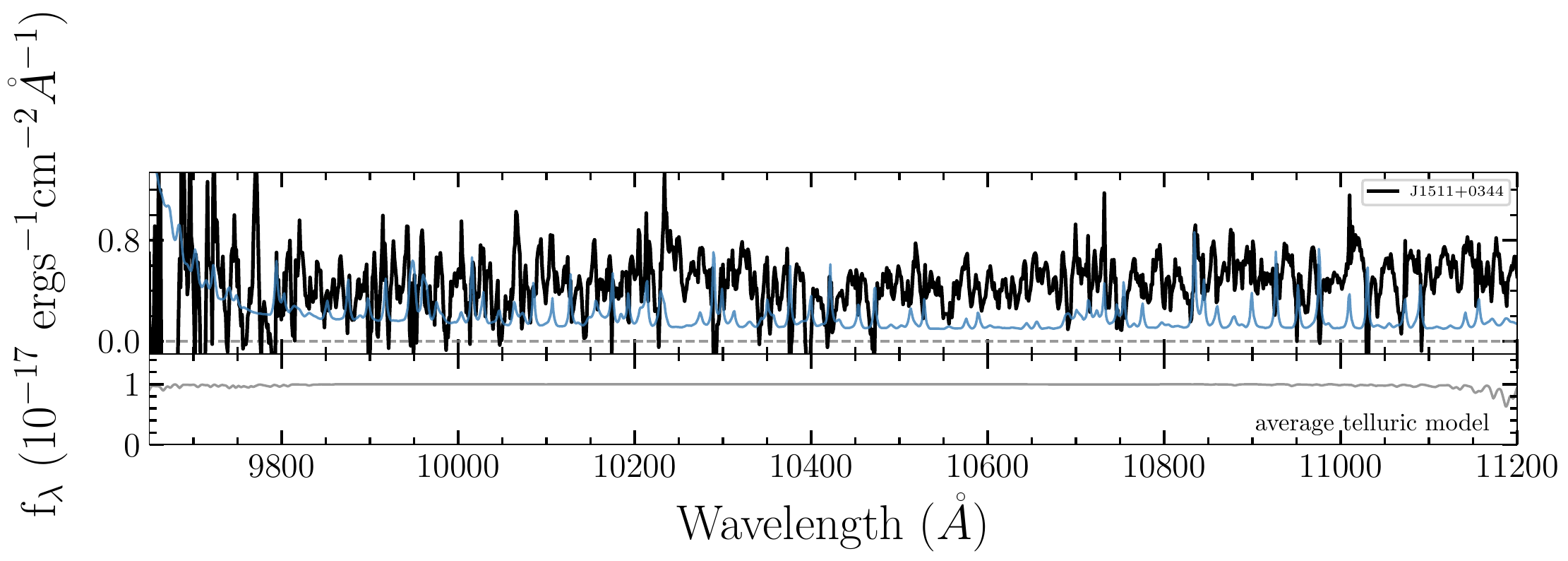}
    \contcaption{Final 1d spectra of UNQ targets in MOSFIRE runs. The 1d spectra here are fluxed and telluric corrected. The blue curves are the noise vectors. The grey curve in the bottom panel is the average telluric model. The spectra and noise vectors are smoothed using the inverse variance with a smoothing window of 5.}
\end{figure*}

\begin{figure*}
    \centering
    \includegraphics[width=\linewidth]{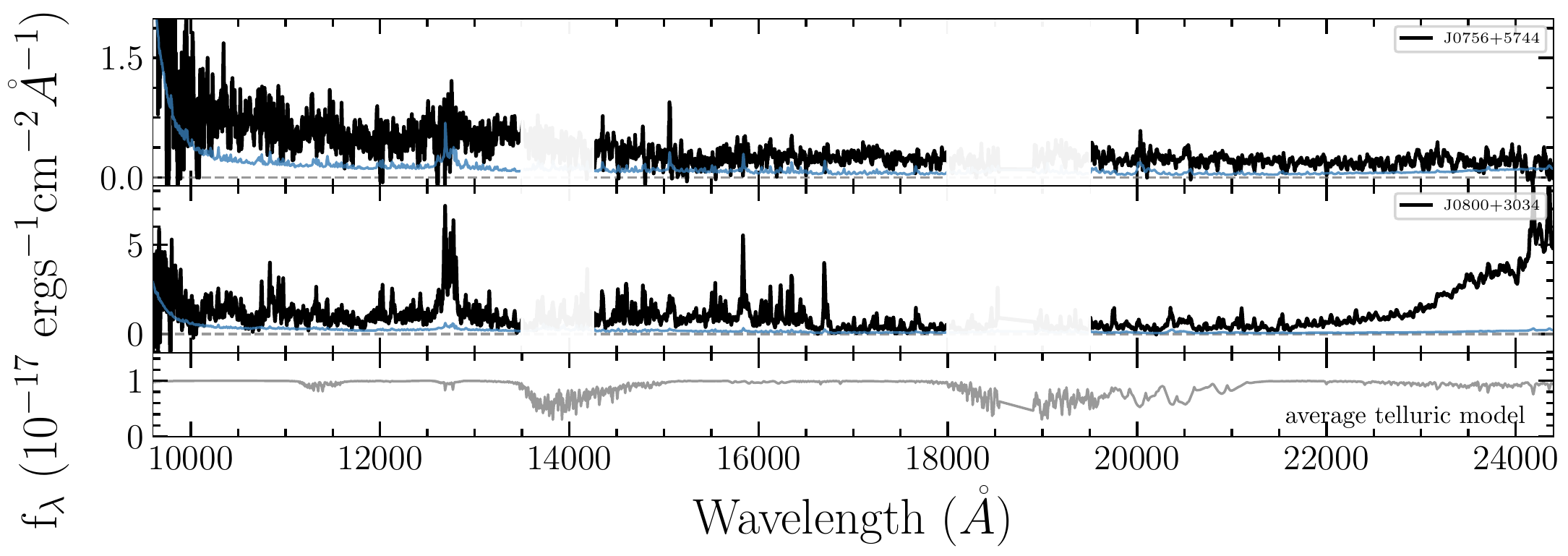}
    \caption{Final 1d spectra of UNQ targets in NIRES runs. The 1d spectra here are fluxed and telluric corrected. The blue curves are the noise vectors. The grey curve in the bottom panel is the average telluric model. The spectra and noise vectors are smoothed using the inverse variance with a smoothing window of 5.}
    \label{fig:nires_uNQ}
\end{figure*}

\bsp	
\label{lastpage}
\end{document}